\documentclass[twocolumn,tighten]{aastex631}
\usepackage{graphicx}
\usepackage{amsmath}
\usepackage{natbib}
\usepackage{amssymb}

\usepackage{sidecap}

\newcommand       \cqg         {CQG}
\newcommand       \jcph        {JCPH}
\shorttitle{Neutrino-Driven Winds from Rotating Protoneutron Stars}
\shortauthors{D.~Desai, D.~M.~Siegel, \& B.~D.~Metzger}

\begin{document}

\newcommand{\be}{\begin{equation}}
\newcommand{\ee}{\end{equation}}

\newcommand  \gcc {~\mathrm{g}~\mathrm{cm}^{-3}}
\newcommand  \E {\times 10^}

\title{Three-Dimensional General-Relativistic Simulations of Neutrino-Driven Winds from Rotating Proto-Neutron Stars}

\author[0000-0002-8914-4259]{Dhruv Desai}
\affil{Department of Physics and Columbia Astrophysics Laboratory, Columbia University, Pupin Hall, New York, NY 10027, USA}
\author[0000-0001-6374-6465]{Daniel M.~Siegel}
\affiliation{Perimeter Institute for Theoretical Physics, Waterloo, Ontario N2L 2Y5, Canada}
\affiliation{Department of Physics, University of Guelph, Guelph, Ontario N1G 2W1, Canada}
\author[0000-0002-4670-7509]{Brian D. Metzger}
\affil{Department of Physics and Columbia Astrophysics Laboratory, Columbia University, Pupin Hall, New York, NY 10027, USA}
\affil{Center for Computational Astrophysics, Flatiron Institute, 162 5th Ave, New York, NY 10010, USA} 

\begin{abstract} We explore the effects of rapid rotation on the properties of neutrino-heated winds from proto-neutron stars (PNS) formed in core-collapse supernovae or neutron-star mergers by means of three-dimensional general-relativistic hydrodynamical simulations with M0 neutrino transport.  We focus on conditions characteristic of a few seconds into the PNS cooling evolution when the neutrino luminosities obey $L_{\nu_e} + L_{\bar{\nu}_e} \approx 7\times 10^{51}$ erg s$^{-1}$, and over which most of the wind mass-loss will occur.
After an initial transient phase, all of our models reach approximately steady-state outflow solutions with positive energies and sonic surfaces captured on the computational grid.  Our non-rotating and slower-rotating models (angular velocity relative to Keplerian $\Omega/\Omega_{\rm K} \lesssim 0.4$; spin period $P \gtrsim 2$ ms) generate approximately spherically symmetric outflows with properties in good agreement with previous PNS wind studies.  By contrast, our most rapidly spinning PNS solutions ($\Omega/\Omega_{\rm K} \gtrsim 0.75$; $P \approx 1$ ms) generate outflows focused in the rotational equatorial plane with much higher mass-loss rates (by over an order of magnitude), lower velocities, lower entropy, and lower asymptotic electron fractions, than otherwise similar non-rotating wind solutions.  Although such rapidly spinning PNS are likely rare in nature, their atypical nucleosynthetic composition and outsized mass yields could render them important contributors of light neutron-rich nuclei compared to more common slowly rotating PNS birth.  Our calculations pave the way to including the combined effects of rotation and a dynamically-important large-scale magnetic field on the wind properties within a 3D GRMHD framework.
\end{abstract}


\section{Introduction}
\label{sec:intro}

The aftermath of a successful core-collapse supernova explosion is the formation of a hot, proto-neutron star (PNS) that cools via the emission of thermal neutrinos over the ensuing seconds, radiating the gravitational binding energy of the star (e.g., \citealt{Burrows&Lattimer86,Pons+99,Roberts12}; see \citealt{Roberts&Reddy17} for a recent review).  These neutrinos deposit energy into the atmosphere of the PNS, driving an outflow of mass known as the {\it neutrino-driven wind} (e.g., \citealt{Duncan+86,Qian&Woosley96,Thompson+01}).  A similar PNS cooling phase, and concomitant neutrino-driven wind, accompanies the cooling evolution of the remnant of a neutron star merger (e.g., \citealt{Dessart+09,metzger_red_2014,perego_neutrino-driven_2014,Kaplan+14,Metzger+18}), in cases when the remnant does not promptly collapse into a black hole.  

The neutrino wind has long been considered a potential site for the nucleosynthesis of heavy neutron-rich isotopes through the rapid neutron capture process ($r$-process; e.g., \citealt{Meyer+92,Takahashi+94,Woosley+94}).  The many past studies of neutrino-driven winds have primarily been focused on spherically symmetric, non-rotating PNS winds accelerated by thermal pressure (e.g., \citealt{Kajino+00,Sumiyoshi+00,Otsuki+00,Thompson+01,Arcones+07,fischer_protoneutron_2010,Roberts+10,Arcones&Montes11,Roberts+12a,MartinezPinedo+12,Fischer+12}).  This body of work has led to the conclusion that normal PNS winds fail to achieve the conditions necessary for nucleosynthesis to reach the third $r$-process peak around an atomic mass number $A \sim 195$. The latter requires an outflow with a combination of high specific entropy $s_{\infty}$, short expansion timescale $\tau_{\rm exp}$, and low electron fraction $Y_e$ \citep{Hoffman+97,Meyer&Brown97} as it passes through the radii where seed nuclei form.  In particular, even for only moderately neutron-rich conditions (e.g., $0.4 \lesssim Y_e \lesssim 0.5$) a sufficiently large value of $s_{\infty}^{3}/(Y_e^3\tau_{\rm exp})$ results in a high ratio of neutrons to seed nuclei$-$and hence a successful heavy $r$-process$-$by trapping protons into $\alpha$-particles as a result of the freeze-out of the neutron-modified triple-$\alpha$ reaction $^{4}$He($\alpha$n,$\gamma$)$^{9}$Be($\alpha$,n)$^{12}$C (e.g., \citealt{Meyer+92,Woosley&Hoffman92}).

Several ideas have been proposed beyond the standard scenario in order to achieve a high neutron-to-seed ratio, and a successful second- or third-peak $r$-process.  These include postulating the existence of additional sources of heating (e.g., damping of convectively-excited waves; \citealt{Suzuki&Nagataki05,Metzger+07,Gossan+20}) or by resorting to extreme parameters, such as massive $\gtrsim 2.2M_{\odot}$ neutron stars \citep{Wanajo13} or those with extremely strong magnetic fields (``magnetars''; \citealt{Thompson03,Thompson+04,Metzger+07,Metzger+08,Vlasov+14,Vlasov+17}).  

Insofar as rapidly spinning magnetars are contenders for the central engines of gamma-ray bursts (e.g., \citealt{Thompson+04,bucciantini_magnetar-driven_2007,Metzger+11a}), the nuclear composition of their outflows may have important implications for the gamma-ray emission mechanism (e.g., \citealt{beloborodov_collisional_2010}) and the composition of cosmic rays accelerated in the relativistic jet (e.g., \citealt{Metzger+11b,Bhattacharya+21}).  Nevertheless, the physical processes responsible for the creation of an ordered large-scale magnetic field during the PNS phase remain uncertain and subject to active research (e.g., \citealt{Raynaud+20}). 

Two of the potentially important ingredients in neutrino-driven winds, which we explore in this work, are the effects of general relativity (GR) and rapid rotation.  The deeper gravitational potential well of the PNS present in GR tends to increase the entropy of the outflows relative to an otherwise equivalent model with Newtonian gravity by around $50\%$ (e.g., \citealt{Cardall&Fuller97,Otsuki+00,Thompson+01}).  Rotation, on the other hand, will generally act to decrease the entropy of the outflows, by reducing the effective gravitational potential due to centrifugal effects (e.g., \citealt{Metzger+07}).  Extremely rapid rotation could in principle also reduce the wind electron fraction, in part because fewer neutrino absorptions per nucleon are necessary to unbind the wind material near the rotational equator, allowing the outflow's composition to remain closer to that of the highly neutron-rich PNS surface (e.g., \citealt{Metzger+08}).

Beyond parametrized one-dimensional models (e.g. \citealt{Duncan+86,Qian&Woosley96,Thompson+01}), numerical work on neutrino-driven winds has focused on 1D and 2D Newtonian hydrodynamical simulations with approximate neutrino transport (e.g., \citealt{Arcones+07,Hudephol+10,fischer_protoneutron_2010,Roberts+10,Nakazato+13,Dessart+09,Arcones+11}). Three-dimensional simulations have so far concentrated on the neutron star merger case, including the Newtonian simulations by \citet{perego_neutrino-driven_2014}. Three-dimensional simulations in the context of core-collapse supernovae have so far focused on the early post-bounce evolution and the explosion mechanism itself, rather than on the long-term cooling evolution of the PNS (e.g., \citealt{Burrows+20}).

In this paper, we explore the effects of rapid rotation on neutrino-heated PNS winds by means of general-relativistic hydrodynamical simulations with approximate neutrino transport. Rather than employing initial conditions for the PNS motivated by self-consistent supernova or merger simulations, we instead follow previous work \citep{Kaplan+14} in constructing parameterized models for the thermodynamic and compositional structure of the PNS that result in neutrino luminosities and energies consistent with those predicted by successful supernova and neutron star merger simulations to occur a few seconds after the birth of the star (the epoch over which most of the integrated wind mass-loss occurs).  By first isolating the effects of rapid rotation in the purely hydrodynamical context, our work here also paves the way for future simulations which will include additional effects, such as the presence of a strong ordered magnetic field.

This paper is organized as follows.  In Section~\ref{sec:methods}, we describe the physical set-up, initial conditions, and numerical code used to perform our neutrino-wind simulations. In Section~\ref{sec:results}, we describe our results, starting with non-rotating PNS wind solutions and then moving on to the rotating cases.  As we shall discuss, rapid ($\sim$ millisecond period) rotation can have large effects on essentially all of the key wind properties.  In Section~\ref{sec:conclusions} we summarize our conclusions and speculate on the potential role of rapidly spinning PNS birth as sources of heavy neutron-rich nuclei.

\section{Methodology}
\label{sec:methods}

\subsection{Numerical Evolution Code}

Our simulations of PNS winds are performed in three-dimensional general-relativistic hydrodynamics (GRHD) using a modified version of \texttt{GRHydro} \citep{Mosta+14} as described in \citet{Siegel&Metzger18}, which is built on the open-source \texttt{Einstein Toolkit}\footnote{\href{http://einsteintoolkit.org}{http://einsteintoolkit.org}}\citep{goodale_cactus_2003,schnetter_evolutions_2004,thornburg_black-hole_2004,Loffler+12,babiuc-hamilton_einstein_toolkit_2019}. This code implements the equations of ideal general-relativistic magnetohydrodynamics with a finite-volume scheme using piecewise parabolic reconstruction \citep{colella_piecewise_1984} and the approximate HLLE Riemann solver \citep{harten_upstream_1983}. Recovery of primitive variables is implemented using the framework presented in \citet{siegel_recovery_2018} and \citet{siegel_grmhd_con2prim_2018}, which provides support for any composition-dependent, three-parameter equation of state (EOS). The magnetic field is evolved using a variant of constrained transport (the ``flux-CT'' method; \citealt{toth_nablacdot_2000}) in order to maintain the solenoidal constraint. In the present study of purely hydrodynamical winds, however, the magnetic field is ignored. Its initial field strength is set to a very small number and monitored throughout the evolution to ensure that it does not impact the dynamics of the simulation.

Although the code is capable of evolving spacetime, the present set of simulations employs a fixed metric for computational efficiency, determined self-consistently from the matter distribution of the initial conditions (Sec.~\ref{sec:models}). A fixed spacetime is a good approximation here, because the star's structure remains nearly constant in time and less than a fraction of $\sim\!10^{-5}$ of the star's mass is removed by winds over the duration of the simulation.  We also performed a test simulation including the full metric evolution, which exhibited only small differences from the fixed-metric case.

We consider both non-rotating and rotating PNS models (Sec.~\ref{sec:models}). The computational domain is set-up as a Cartesian grid hierarchy consisting of one base grid and six nested refinement levels for our non-rotating PNS model.  In our fiducial non-rotating PNS model $\texttt{nrot-HR}$, the finest and smallest grid is a $15 \times 15 \times 15$\,km box centered at the origin and the center of the star, with a resolution of $\Delta x \simeq 225$ m. The size of the largest box is $960\,\mathrm{km} \times 960\,\mathrm{km} \times 960\,\mathrm{km}$, which allows us to capture the wind zone and to determine the asymptotic properties of the wind.  In comparison, the grid setup for our most rapidly rotating PNS models (\texttt{rot.7-MR}, \texttt{rot.6-MR}) has one less refinement level, with the finest and smallest grid being a $30 \times 30 \times 30$\ km box; this is necessary to resolve the high-velocity outflows from the PNS surface out to larger radii.  All of our rotating models employ a spatial resolution on the finest grid of $\Delta x \simeq 450$ m.

\begin{figure}
    \includegraphics[width=0.49\textwidth]{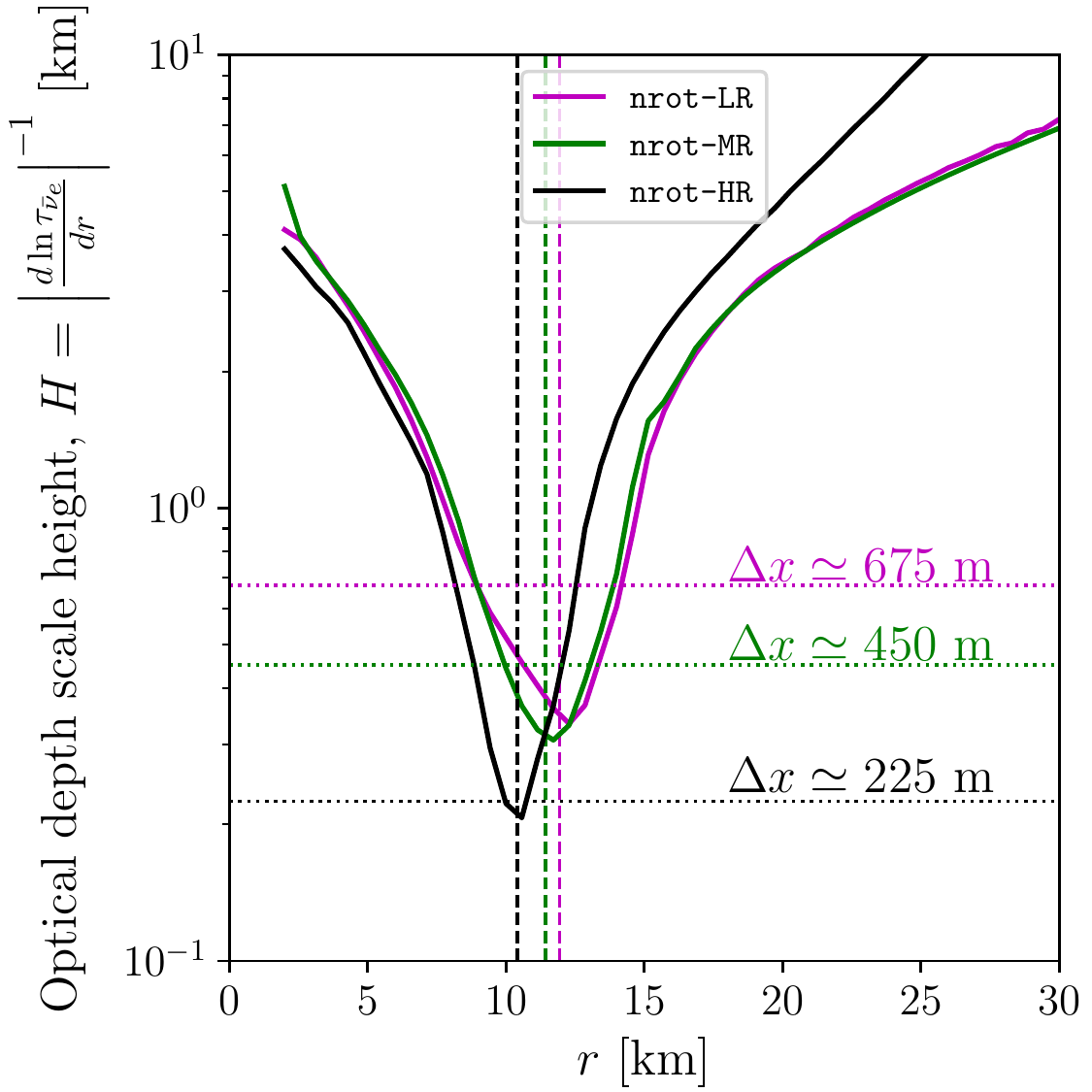}
    \caption{Scale height $H=\left|\frac{d \ln \tau_{\bar \nu_e}}{dr}\right|^{-1}$ associated with the $\bar{\nu}_e$ optical depth $\tau_{\bar{\nu}_e}$, as a function of radius $r$ from the center of the PNS (solid lines), averaged over polar angle ($\theta = 0^\circ - 90^\circ$) and time-averaged from 50 to 100 ms, as well as the resolution $\Delta x$ of the innermost grid (dotted lines), for our three non-rotating PNS solutions with different spatial resolution:  $\texttt{nrot-MR}$ (black, $\Delta x \simeq 450$ m), $\texttt{nrot-LR}$ (purple, $\Delta x \simeq 675$ m), and $\texttt{nrot-HR}$ (green, $\Delta x \simeq 225$ m).  None of the simulations resolve the region around the neutrinosphere radius ($\tau_{\bar \nu_e}=1$; vertical dashed lines) with several grid points, and hence cannot accurately converge on the neutrino luminosity or mean energy.  However, the gain region of net heating and wind zone on larger scales is well resolved by even the lowest resolution runs.}
\label{fig:scale_height}
\end{figure}

We find that our adopted resolution is not sufficient to resolve the neutrino decoupling region near the PNS surface, which we quantify in Fig.~\ref{fig:scale_height} using the optical depth scale-height near the neutrinosphere obtained once our wind solutions have reached a steady state.  This would require a resolution that is approximately a factor of 10 higher than the highest resolution run we have explored (\texttt{nrot-HR}, with $\Delta x \simeq 225$ m), which is computationally infeasible for this study.  As discussed below, we therefore do not (nor would we expect to) obtain convergent values for the steady-state neutrino luminosities or mean energies (Figs.~\ref{fig:conv_evol} and \ref{fig:rot_evol}), as these are determined near the decoupling surface.  However, this deficiency is not critical for the purposes of this study, because the neutrino radiation field serves primarily as a boundary condition controlling the wind heating and compositional changes at larger radii above the neutrinosphere$-$regions which are properly resolved.  For example, our fiducial simulation $\texttt{nrot-HR}$ resolves the temperature scale-height in the decoupling region with at least 4-5 points while obtaining reasonably convergent hydrodynamic wind properties.  Furthermore, as discussed in Section~\ref{sec:results}, the key wind properties (e.g., mass loss rate, entropy) obtained by our non-rotating PNS simulations performed at lower spatial resolution than the fiducial model ($\texttt{nrot-MR}$ and $\texttt{nrot-LR}$, respectively) do agree with one another to $\lesssim 10\%$, once the impact of their different neutrino luminosities and energies are accounted for as predicted by analytic scaling relations \citep{qian_nucleosynthesis_1996}.  

In order to reduce computational costs, we make use of appropriate symmetries.  For our rotating models, the $z$-axis corresponds to the rotational axis; thus for all models we employ $180^\circ$ rotational symmetry around the $z$-axis, as well as reflection symmetry across the $xy$-plane. To ensure that our results do not depend on the choice of symmetry, we performed one of our rotating simulations with and without imposing symmetry, finding indistinguishable results between the two cases.

Weak interactions and approximate neutrino transport are included via a leakage scheme \citep{bruenn_stellar_1985,ruffert_coalescing_1996} following the implementation of \citet{galeazzi_implementation_2013} and \citet{radice_dynamical_2016} as described in \citet{Siegel&Metzger18}, together with a `ray-by-ray' transport scheme (`M0' scheme; \citealt{radice_dynamical_2016,radice_binary_2018}).  The M0 scheme represents an approximation to neutrino transport derived by taking the first moment of the Boltzmann equation and using a closure relation that assumes neutrinos stream along radial rays at the speed of light. Neutrino mean energies and number densities are evolved according to radial evolution equations (Eqs.~(A11) and (A15) of \citealt{radice_dynamical_2016}). This ray-by-ray transport scheme does not account for potential neutrino interactions between different rays (i.e.~lateral transport). The M0 transport scheme will be least accurate in regions of high optical depth, i.e.~within the PNS for our study.  As a result, the neutrino properties within the PNS may not be fully consistent with the temperature profile. However, in this region we initiate a somewhat {\it ad hoc} radial temperature profile (Sect.~\ref{sec:models}), rather than one based on a self-consistent evolution (e.g., from the stellar core collapse or neutron star merger).  In the spirit of previous work (e.g., \citealt{Thompson+01}) our primary goal is instead to characterize the properties of the resulting wind at a given neutrino luminosity and mean energy.  Thus, the radial approximation should be sufficient because the wind is launched above the surface of the PNS where the optical depth is low, and is furthermore roughly spherically symmetric, even in the fastest rotating models.

 The leakage scheme includes charged-current $\beta$-processes, electron-positron pair annihilation, and plasmon decay. Neutrino opacities include neutrino absorption by nucleons and coherent scattering on free nucleons and heavy nuclei. We neglect the effects of magnetic fields on the neutrino opacities, which is a good approximation except unless the PNS is extremely highly magnetized (e.g., \citealt{beloborodov_nuclear_2003,Duan&Qian04}). Optical depths are calculated using the quasi-local scheme presented by \citet{neilsen_magnetized_2014}.

The neutrino evolution is coupled to the equations of GRHD in an operator-split fashion, leading to electron fraction changes as well as neutrino energy and momentum deposition to matter. The dominant weak reactions for heating the wind and changing the electron fraction are
\be
\nu_e + n \longleftrightarrow e^- + p~~~\mathrm{and}~~~ \bar \nu_e + p \longleftrightarrow e^+ + n,
\label{eq:ntop}
\ee
where $\nu_e$ and $\bar \nu_e$ are the electron neutrino and anti-neutrino, $p$ is the proton, $n$ is the neutron, and $e^-$ and $e^+$ are the electron and positron, respectively.

Neutrino energies and number densities are evolved on radial rays that represent a uniform spherical `M0 grid' extending radially to 200\,km, with $n_r \times n_\theta \times n_\phi = 600 \times 20 \times 40$ grid points. As neutrino transport quantities only slowly change in time with respect to the hydro time step, and for computational efficiency, effective neutrino absorption is updated via M0 only every 16 time steps of the hydro evolution. Neutrino transport includes three neutrino species: electron neutrinos and anti-neutrinos ($\nu_e, \bar \nu_e$), as well as all $\mu$ and $\tau$ neutrinos and anti-neutrinos grouped into one additional category ($\nu_x$). The grid covers the neutrinosphere (for all species), defined by the surface above which the neutrino optical depth to infinity lies between 0.7 and 1, and the gain layer, where net neutrino heating unbinds matter from the PNS surface. The grid also extends to large enough radii to cover all densities for which weak interactions are expected to be appreciable ($\rho \gtrsim 10^{4} \gcc$). 

\subsection{Neutron Star Models and Initial Conditions}
\label{sec:models}

For initial conditions, we construct axisymmetric hydrostatic profiles of non-rotating and solid-body rotating neutron stars of gravitational mass $1.4 M_{\odot}$ with the \texttt{RNS} code \citep{Stergioulas+95}. Our grid of simulations, and the key properties of the PNS for each model, are summarized in Table \ref{tab:models}. 

For the equation of state (EOS), we adopt the SFHo\footnote{Available in tabulated form on \hyperlink{stellarcollapse.org}{stellarcollapse.org}.} model \citep{Steiner+13}, which covers densities from $10^{-12}$ to 10 fm$^{-3}$ and temperatures from 0.1 to 160 MeV.  The EOS parameters are calibrated to nuclear binding energies as well as to other observational and experimental constraints. For densities below saturation density, the model accounts for light and heavy nuclei formation, and smoothly transitions from nuclei to uniform nuclear matter with a thermodynamically consistent excluded volume description \citep{Hempel+10}.  A distribution of different nuclear species are assumed, rather than just the single nuclei approximation. Results for light nuclei are in agreement with quantum many-body models. The presence of nuclei ensures that nuclear binding energy released as individual nucleons recombine into light nuclei is captured by the flow; conversion of this energy into kinetic energy can significantly affect the unbinding of winds (cf.~Sec.~\ref{sec:results}). For this EOS, the radius of a $1.4 M_{\odot}$ non-rotating cold neutron star is 11.88 km and the maximum stable mass is 2.059 $M_{\odot}$. 

The initial temperature profile of the star as a function of density, $T(\rho)$, must be specified as an initial condition.  Although our goal is to study PNS winds generated following a core-collapse supernova (or, potentially, a neutron star merger), we do not obtain the initial temperature profiles directly from supernova or merger simulations.  Rather, we specify $T(\rho)$ as an {\it ad hoc} functional form following \citet{Kaplan+14}, the parameters of which are so chosen to generate steady-state neutrino emission properties from the star similar to those predicted a few seconds after a successful core collapse explosion (e.g., \citealt{Pons+99}) or neutron star merger event (e.g., \citealt{Dessart+09}).  Specifically, we adopt a temperature profile that smoothly transitions from the hot PNS core temperature $T_{\mathrm{max}}$ to a colder atmosphere temperature $T_{\mathrm{min}}$ (Eq.~(A1) of \citealt{Kaplan+14}):
\be
T(\rho) = T_{\rm min} + \frac{T_{\rm max}}{2}\left( \tanh \frac{\left( \log_{10}(\rho)-\bar{m}\right)}{\bar{s}}+1\right),
\ee
where $\bar{m}$ is the midpoint of the logarithmic density roll-off and $\bar{s}$ is the e-folding scale.  We use the temperature profile denoted \texttt{C20p0} in \citet{Kaplan+14}, with parameters set as follows: $\bar{m} = 14.2$, $\bar{s} = 0.3$, $T_{\mathrm{min}}= 0.015$ MeV, and $T_{\mathrm{max}} = 20$ MeV. These parameter values were chosen to give rise to approximate target values of the neutrino luminosities and mean energies, once a steady wind has been established.

As with the temperature profile, the initial electron fraction profile $Y_e(\rho, T)$ must be specified. Given the EOS and temperature profile, we again follow \citet{Kaplan+14} and set
\begin{eqnarray}
Y_{e}(\rho, T[\rho]) &=& Y_{e,\nu-\mathrm{free} \beta}(\rho) (1-e^{-\rho_{\mathrm{trap}/}\rho}) \nonumber \\
&+& Y_{e,\nu-\mathrm{trap} \beta}(\rho, T[\rho]) e^{-\rho_{\mathrm{trap}}/\rho}, \label{eq:Ye_profile}
\end{eqnarray}
where $Y_{e,\nu-\mathrm{free} \beta}$ denotes the electron fraction for cold matter in $\beta$-equilibrium without neutrinos (computed at $T_{\rm min}$)  and $Y_{e,\nu-\mathrm{trap} \beta}$ refers to the electron fraction for hot and dense matter in $\beta$-equilibrium with neutrinos present. 
For densities $\rho \ll \rho_{\rm trap} \sim 10^{12.5} \gcc$, neutrinos decouple from matter. To account for free streaming neutrinos near the PNS surface ($\rho \lesssim \rho_{\mathrm{trap}}$), the attenuation factor $e^{-\rho_{\mathrm{trap}}/\rho}$ serves to smoothly connect the hot and cold matter solutions.

The function $Y_{e,\nu-\mathrm{free} \beta}$ in Eq.~\eqref{eq:Ye_profile} is obtained from the EOS according to the condition
\be
\mu_{\nu} = 0 = \mu_p - \mu_n + \mu_e,
\label{eq:munu}
\ee
at $T_{\rm min}$, where $\mu_{\nu}$, $\mu_n$, $\mu_p$, $\mu_e$ are the chemical potentials of the neutrinos, neutrons, protons, and electrons, respectively. As the neutrino density $n_\nu$ is negligible in the free-streaming regions, the lepton fraction (ratio of lepton to baryon number) obeys $Y_{\mathrm{lep}} = Y_{e,\nu-\mathrm{free} \beta}$.

We calculate $Y_{e,\nu-\mathrm{trap} \beta}$ by treating neutrinos as a relativistic Fermi gas in equilibrium, computing the neutrino fraction $Y_\nu$ according to
\be
Y_\nu = \frac{n_\nu}{\rho N_A},
\ee
where $n_\nu$ is obtained from Fermi integral relations (Eq.~(B5) of \citealt{Kaplan+14}). We then iteratively solve the relation
\be
0 = Y_{\mathrm{lep}} - (Y_{e,\nu-\mathrm{trap}\beta} +Y_\nu),
\ee
where $Y_{\mathrm{lep}}$ is computed from Eq.~\eqref{eq:munu}, used here as a fixed input.  The value of $Y_e$ dictated by this prescription is roughly constant for $\rho \lesssim 10^7 \gcc$ (corresponding to the ``atmosphere'' on the simulation grid), so we initialize $Y_e(\rho \lesssim 3\times 10^{6} \gcc$) to $Y_{\mathrm{atm}} \approx 0.46$.  Likewise, we set the density and temperature of the initial atmosphere to $\rho_{\mathrm{atm}}=340 \gcc$ and $T_{\mathrm{atm}}=0.015$ MeV, respectively.

Stellar models are computed using \texttt{RNS}, which solves the general-relativistic Euler equations for a uniformly rotating star in axisymmetric spacetime \citep{Stergioulas+95}. In specifying the EOS, \texttt{RNS} requires a table of the energy density as a function of pressure.  We generate this table over the relevant range of densities using the EOS with the temperature and $Y_e$ prescriptions from above.  A stellar model is constructed by specifying a central density $\rho_c$ and a polar to equatorial radius axis ratio $R_p/R_e$. For solutions with arbitrary rotation frequency, \texttt{RNS} first integrates the TOV equations and finds a nearest solution.  The code then estimates solutions through an iterative procedure until the desired ratio $R_p/R_e$ is achieved.  For the non-rotating models, the radius of the neutron star is found to be $R_e=R_p =12.70$ km, larger than the equivalent cold radius of 11.88 km and consistent with the temperature-dependent PNS radius found by \citet{Kaplan+14}.

For our most rapidly rotating model \texttt{rot.6-MR}, we adopt a PNS rotating near-break up, with $\Omega/\Omega_K \approx 0.944$, where $\Omega_K$ is the Keplerian orbital frequency, corresponding to an axis ratio $R_p/R_e=0.6$ ($R_{\rm e} \simeq 17.6$ km) and spin period $P \approx$ 1.11 ms. We also run two cases of intermediate rotation, $R_p/R_e=0.7$ (\texttt{rot.7-MR}; $P = 1.15$ ms) and $R_p/R_e=0.9$ (\texttt{rot.9-MR}; $P = 1.78$ ms).
Further details on the models are given in Table~\ref{tab:models}. 




\begin{table*}
  \begin{center}
    \caption{Suite of 1.4$M_{\odot}$ PNS Simulations}
    \begin{tabular}{c|c|c|c|c|c|c|c|c|c|c}
    \hline
     Model & $R_e^{(a)}$ & $P^{(b)}$ & $R_p/R_e^{(c)}$ & $T/|W|^{(d)}$ & $\Omega/\Omega_{\rm K}^{(e)}$  & $R_{\bar \nu_e}^{(f)}$ & $\langle L_{\nu_e} \rangle^{(g)}$ & $\langle L_{\bar{\nu}_e} \rangle^{(h)}$ & $\langle E_{\nu_e}\rangle^{(i)}$ & $\langle  E_{\bar{\nu}_e} \rangle^{(j)}$ \\
      
- &(km) & (ms) & - & - & - & (km) & (erg s$^{-1}$) & (erg s$^{-1}$) & (MeV) & (MeV)\\
      \hline \hline
\texttt{nrot-LR} & 12.7 & n/a & 1 & 0 & 0 & 12.1 & 5.3e51 & 8.5e51 & 12.5 & 18.0\\

\texttt{nrot-MR} & 12.7 & n/a & 1 & 0 & 0 & 11.4 & 4.1e51 & 6.7e51 & 13.3 & 17.8\\

\texttt{nrot-HR$^*$} & 12.7 & n/a & 1 & 0 & 0 & 10.5 & 2.5e51 & 4.2e51 & 12.8 & 18.3\\
\hline
\texttt{rot.9-MR} & 13.4 & 1.78 & 0.9 & 2.6e-2 & 0.40 & 10.5--11.7 & 2.9e51 & 4.4e51 & 12.5 & 18.4 \\
\hline
\texttt{rot.7-MR}& 15.7 & 1.15 & 0.7 & 7.8e-2 & 0.75 & 9.8--16.4 & 3.3e51 & 5.2e51 & 12.0 & 17.1 \\
\hline
\texttt{rot.6-MR} & 17.6 & 1.11 & 0.6 & 9.4e-2 & 0.94 & 10.0--17.5 & 3.0e51 & 4.9e51 & 11.7 & 17.0 \\
\hline

    \end{tabular}\\
    ${(a)}$ Initial equatorial radius. 
    ${(b)}$ Spin period.
    ${(c)}$ Initial ratio of polar to equatorial radius. 
    ${(d)}$ Ratio of total rotational energy to gravitational binding energy.
    ${(e)}$ Ratio of rotational frequency to Keplerian frequency at stellar equator, $\Omega_{\rm K}= \sqrt{GM/R_e^3}$. 
    ${(f)}$ Steady-state radius of the anti-electron neutrinosphere (or range of radii, in rotating cases).
    ${(g)-(j)}$ Luminosities and mean energies of electron neutrinos and anti-neutrinos, averaged over the final factor of three in simulation time. Model names ending in \texttt{HR, MR, LR} have smallest box resolutions of $225, 450, 675$m, respectively.
    
    $^*$Fiducial non-rotating wind model shown in Figs.~\ref{fig:sph_snaps}, \ref{fig:rad_profs}.
    
\label{tab:models}
  \end{center}
\end{table*}

\subsection{Conditions for $r$-process nucleosynthesis}
\label{sec:nucleosynthesis}

Before describing the results of our simulations, we briefly review the physical processes in the PNS wind that determine whether a successful $r$-process can take place. The surface of the PNS and the inner regions of the wind are sufficiently hot that protons and neutrons exist as free nucleons.  Free nuclei recombine into $\alpha$-particles at radii in the wind where the temperature decreases to $T \lesssim 5\times 10^{9}$ K, as typically occurs $\sim 50-100$ km above the PNS surface in our models (cf.~Figs.~\ref{fig:sph_snaps} and \ref{fig:rot_snaps}). Heavier elements then begin to form as the temperature decreases further, starting with the reaction $^{4}$He($\alpha$n,$\gamma$)$^{9}$Be($\alpha$,n)$^{12}$C for $Y_e \lesssim 0.5$. After $^{12}$C forms, additional $\alpha$ particle captures produce heavy seed nuclei with characteristic mass $\bar{A} \approx 90-120$ and charge $\bar{Z}$ (``alpha-process''; e.g., \citealt{Woosley&Hoffman92}). Finally, the $r$-process itself occurs, as the remaining free neutrons (if any) capture onto these seed nuclei.  

The maximum atomic mass $A_{\rm max}$ to which the $r$-process can proceed depends on the ratio of free neutrons to seed nuclei following the completion of the $\alpha$-process.  Because $^{12}$C-formation is the rate-limiting step to forming the seeds, the neutron to seed ratio (and hence $A_{\rm max}$) depends on the electron fraction $Y_e$, asymptotic wind entropy $s_{\infty}$, and expansion time $\tau_{\rm exp}$ through the seed formation region \citep{Meyer&Brown97,Hoffman+97}.  We follow \citet{Hoffman+97} in defining the latter as\footnote{
\citet{Thompson+01} define a similar quantity, but in terms of the gradient of the density profile rather than the temperature ($\tau_{\mathrm{exp},\rho}$). Given that entropy obeys $s \propto T^{3}/\rho$ for a radiation-dominated plasma and that $s$ is roughly constant in the wind outside the gain region, it follows that $\tau_{\rm exp} \approx 3~\tau_{\mathrm{exp},\rho}$.}
\be
\tau_{\rm exp} \equiv \frac{1}{v^{r}}\left|\frac{d{\rm ln}T}{dr}\right|^{-1}_{\rm T = 0.5{\rm MeV}}.
\label{eq:texp}
\ee
For $\bar{Z}/\bar{A} \approx 0.35-0.4 \lesssim Y_e \lesssim 0.5$, the condition for the $r$-process to reach the second or third peak can be expressed as\footnote{The combination $s^{3}/\tau_{\rm exp}$ enters because the abundance of $^{12}$C nuclei (and hence the number of seed nuclei) created in the wind is equal to an integral of the effective 4-body $^{4}$He($\alpha$n,$\gamma$)$^{9}$Be($\alpha$,n)$^{12}$C reaction rate $\propto \rho^{3}$ times the timescale available for formation $\propto \tau_{\rm dyn};$ for radiation-dominated conditions $s \propto T^{3}/\rho$ and so $\rho \propto s^{-1}$ given that seed nuclei form at roughly a fixed temperature.}
\be
\eta \equiv \frac{s_{\infty}^{3}}{Y_e^{3}\tau_{\rm exp}} \gtrsim \eta_{\rm thr} \approx 
\begin{cases}
4\E{9} & \text{if } A_{\rm max}\sim 135 \text{ (2nd peak)} \cr
9\E{9} & \text{if } A_{\rm max}\sim 195 \text{ (3rd peak)}
\end{cases},
\label{eq:eta}
\ee
where $s_{\infty}$ is expressed\footnote{Throughout the remainder of the paper, specific entropy is expressed in units of $k_B~ \rm baryon^{-1}$ for notational brevity.} in $k_b$ baryon$^{-1}$ and $\tau_{\rm exp}$ in seconds.  Thus, the ratio $\eta/\eta_{\rm thr}$ serves as a ``figure of merit'' for the potential success of a given $r$-process site in the $0.4 \lesssim Y_e \lesssim 0.5$ regime.

Previous studies of the $r$-process in spherical non-rotating PNS winds typically find $\eta \ll \eta_{\rm thr}$, thus disfavoring these events as sources of heavy $r$-process nuclei unless $Y_e \ll 0.4$.  Furthermore, if $Y_e > 0.5$, as suggested by some recent cooling calculations of non-rotating PNS (e.g., \citealt{Pascal+22}), then an $r$-process will not be achieved for any value of $\eta$ (however, see \citealt{Meyer02} for an exception).  

On the other hand, if $Y_e < \bar{Z}/\bar{A} \sim 0.35-0.4$, then an $r$-process is possible for $\eta \ll \eta_{\rm thr}$, with third-peak element production occurring for $Y_e \lesssim 0.25$ \citep{lippuner_r-process_2015}.  The latter is the regime encountered in the dynamical and disk-wind ejecta of neutron star mergers (e.g., \citealt{freiburghaus_r-process_1999,siegel_three-dimensional_2017}) and, potentially, the winds from rapidly spinning proto-magnetars (e.g., \citealt{Metzger+07,Metzger+18}).

\section{Results}
\label{sec:results}
\subsection{Non-Rotating PNS Wind}

\begin{figure*}
    \includegraphics[width=0.49\textwidth]{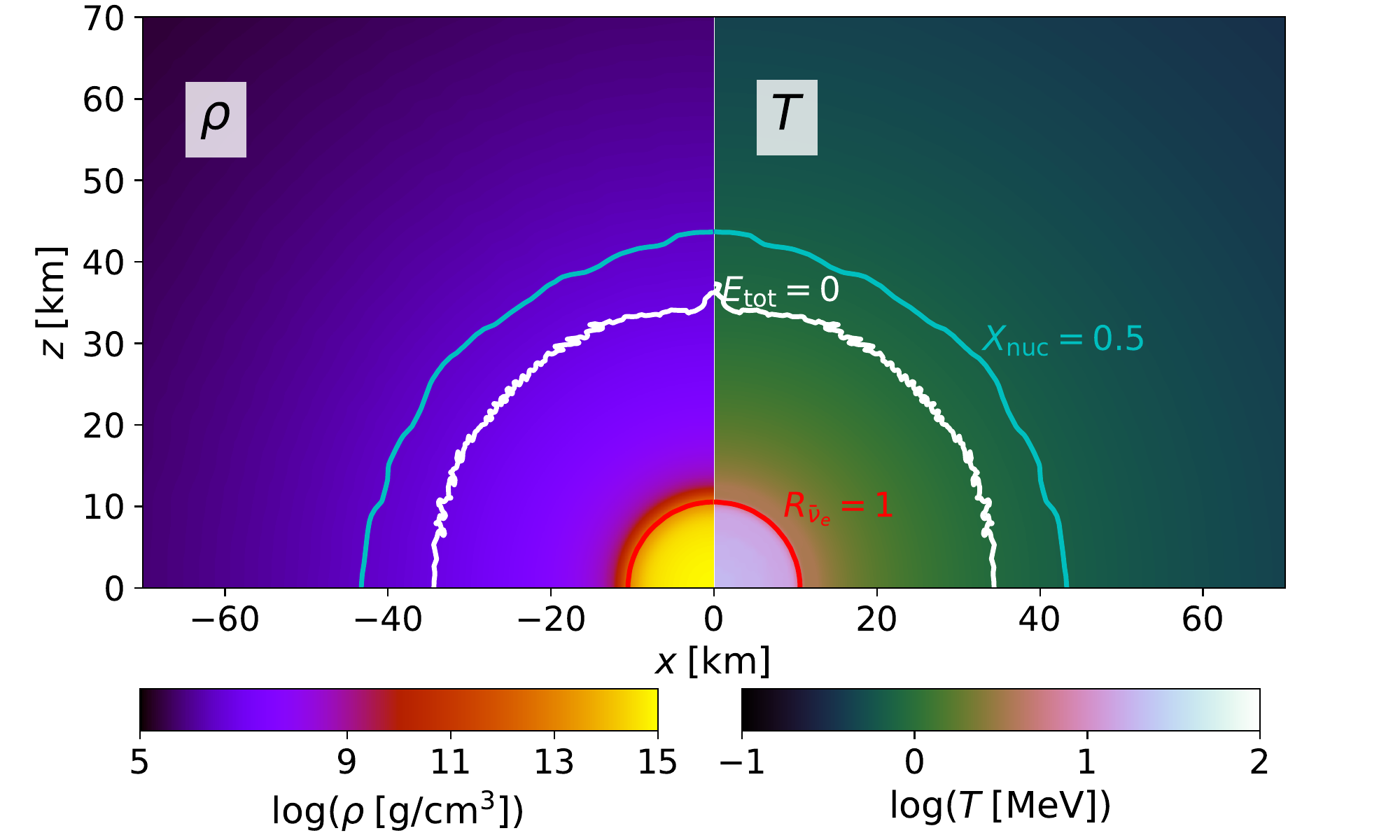}
    \includegraphics[width=0.49\textwidth]{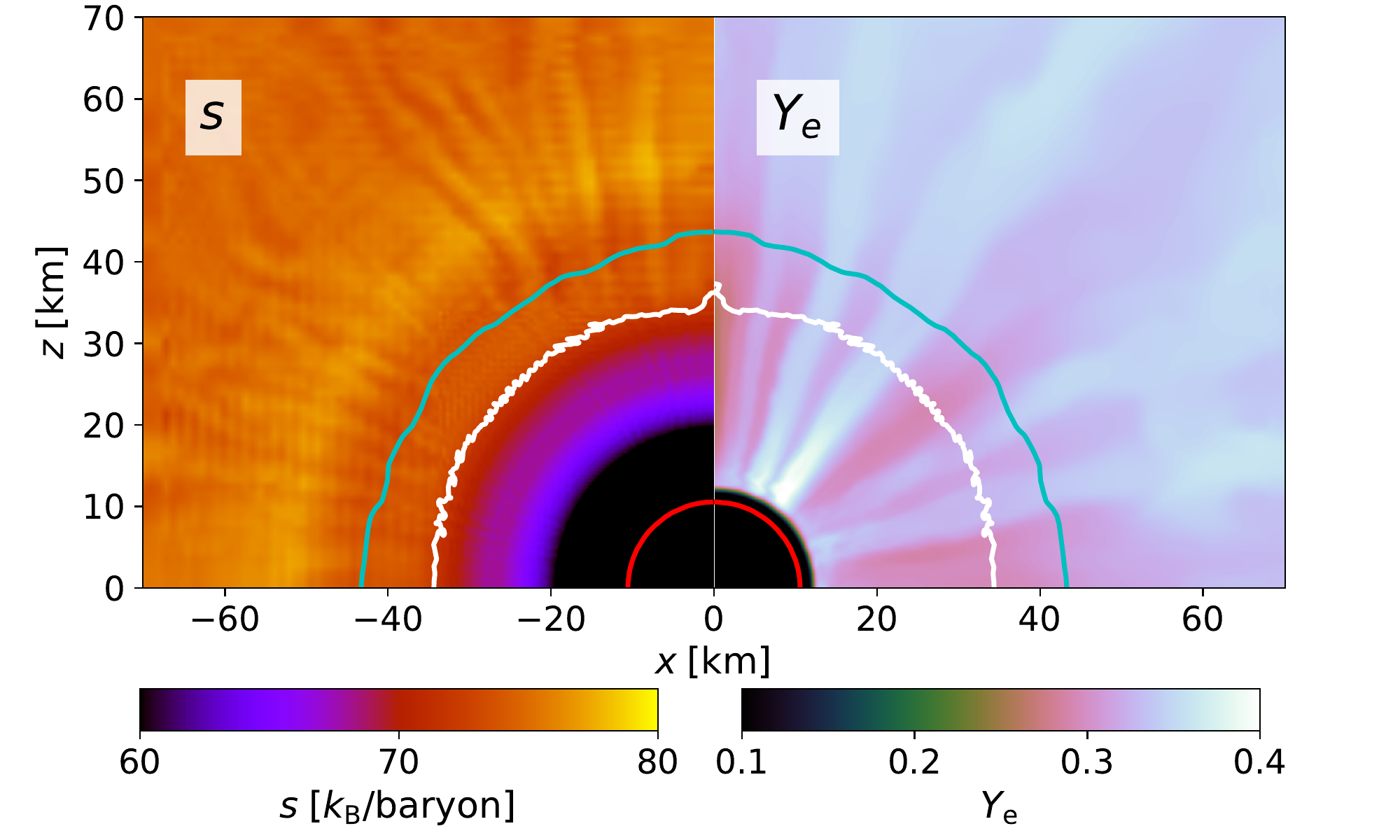}
    \includegraphics[width=0.49\textwidth]{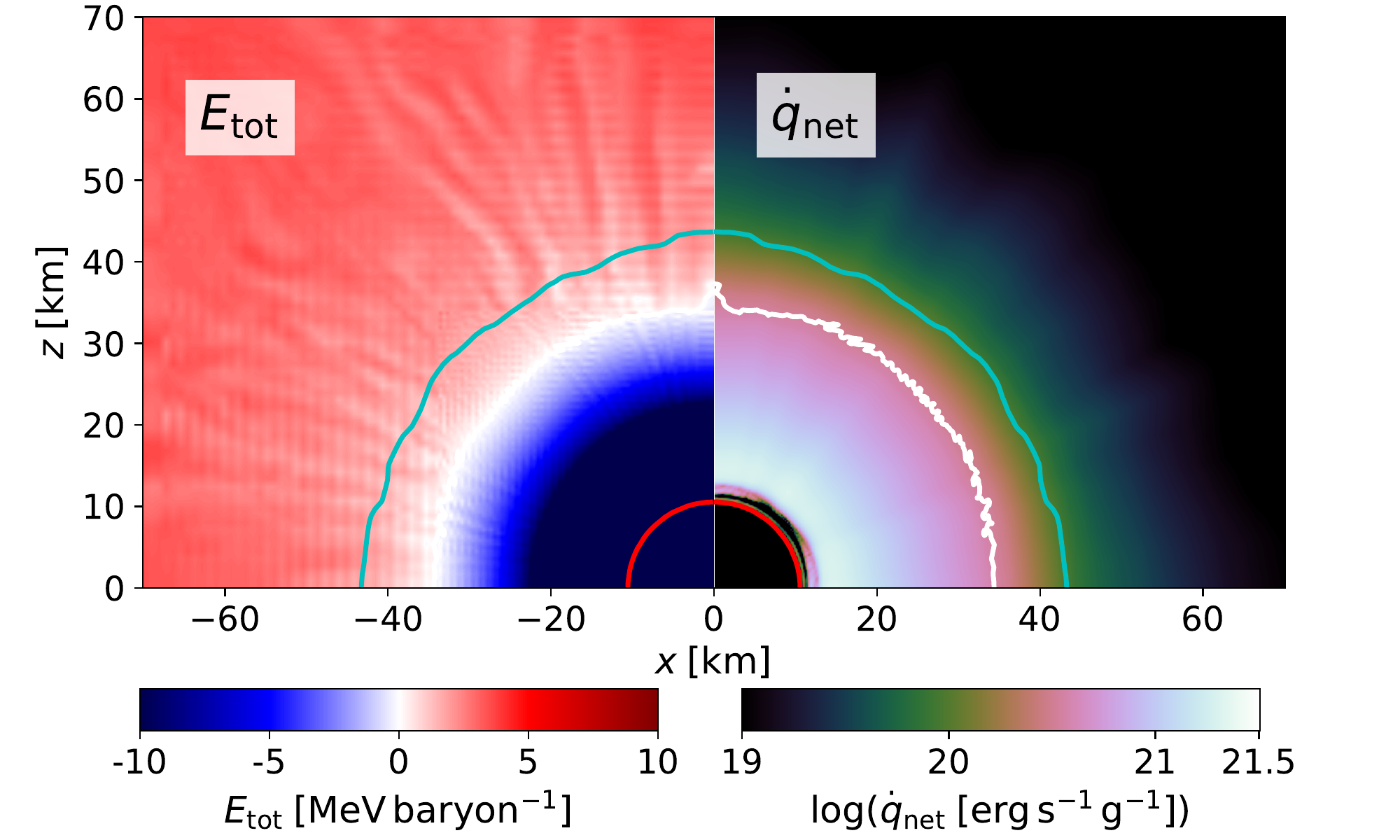}
    \includegraphics[width=0.49\textwidth]{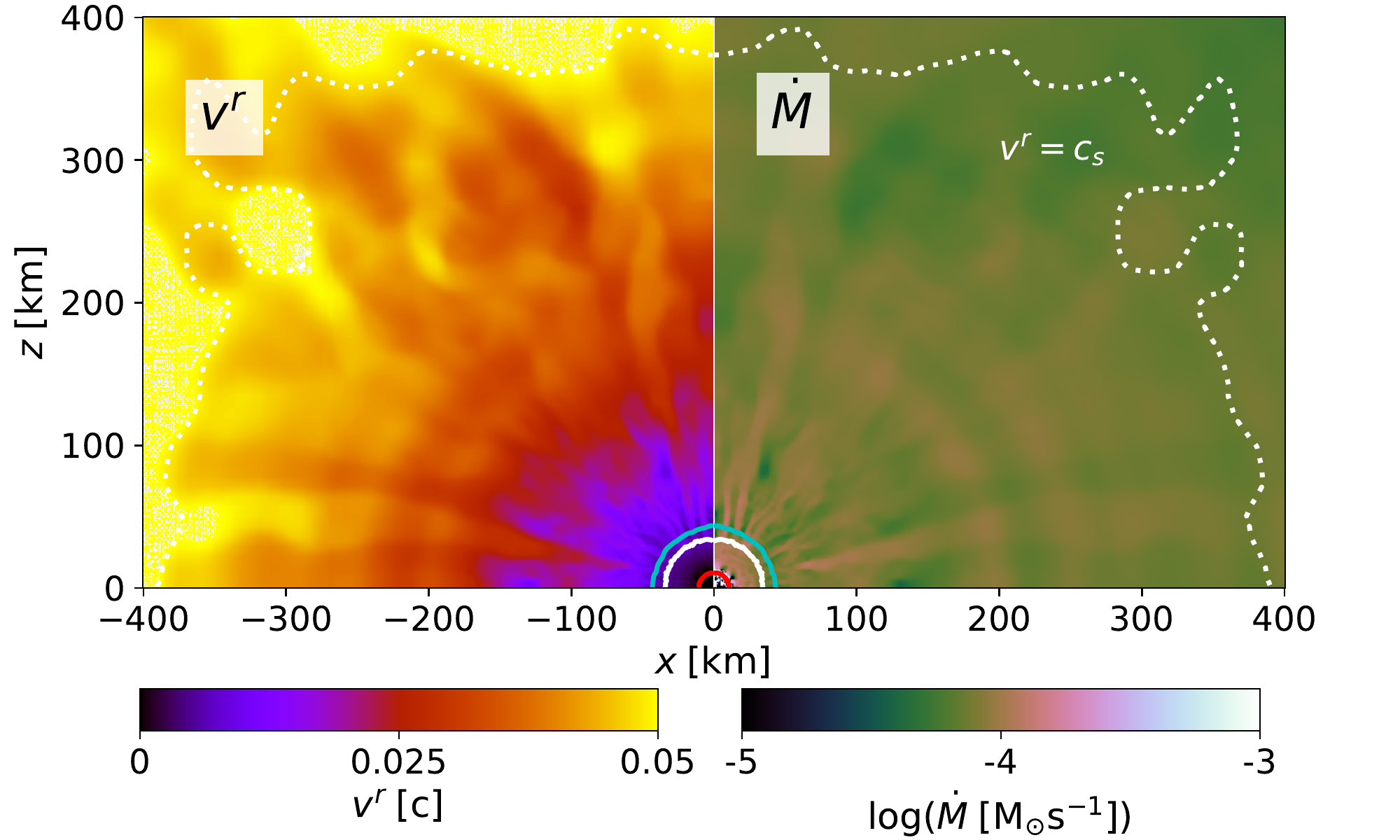}
    \caption{Snapshots of various quantities from our fiducial non-rotating model \texttt{nrot-HR} showing a slice through the $y = 0$ plane at $t = 152$ ms, with contours at the neutrinosphere ($\tau_{\bar \nu_e}=1$; {\it red}), $E_{\rm tot}=0$ ({\it white}), the $\alpha$-particle formation surface ({\it teal}), and the sonic surface (where $v^{r}=c_s$; {\it white dotted}).  The top left panel shows the density $\rho$ and temperature $T$. The top right panel shows the specific entropy $s$ and the electron fraction $Y_e$.
    The bottom left panel shows $E_{\rm tot}$, the total specific energy of wind matter as measured at infinity ($E_{\rm tot}= -h u_t-1$, where $h$ is the specific enthalpy and $u_t$ is the time component of the four-velocity), and the net neutrino heating rate $\dot q_{\rm net}$. The bottom right panel shows the radial velocity $v^r$ and the isotropic-equivalent mass loss rate $\dot M = 4\pi r^{2} \rho v^r$. 
    }
\label{fig:sph_snaps}
\end{figure*}

\begin{figure}
    \includegraphics*[width=0.48\textwidth]{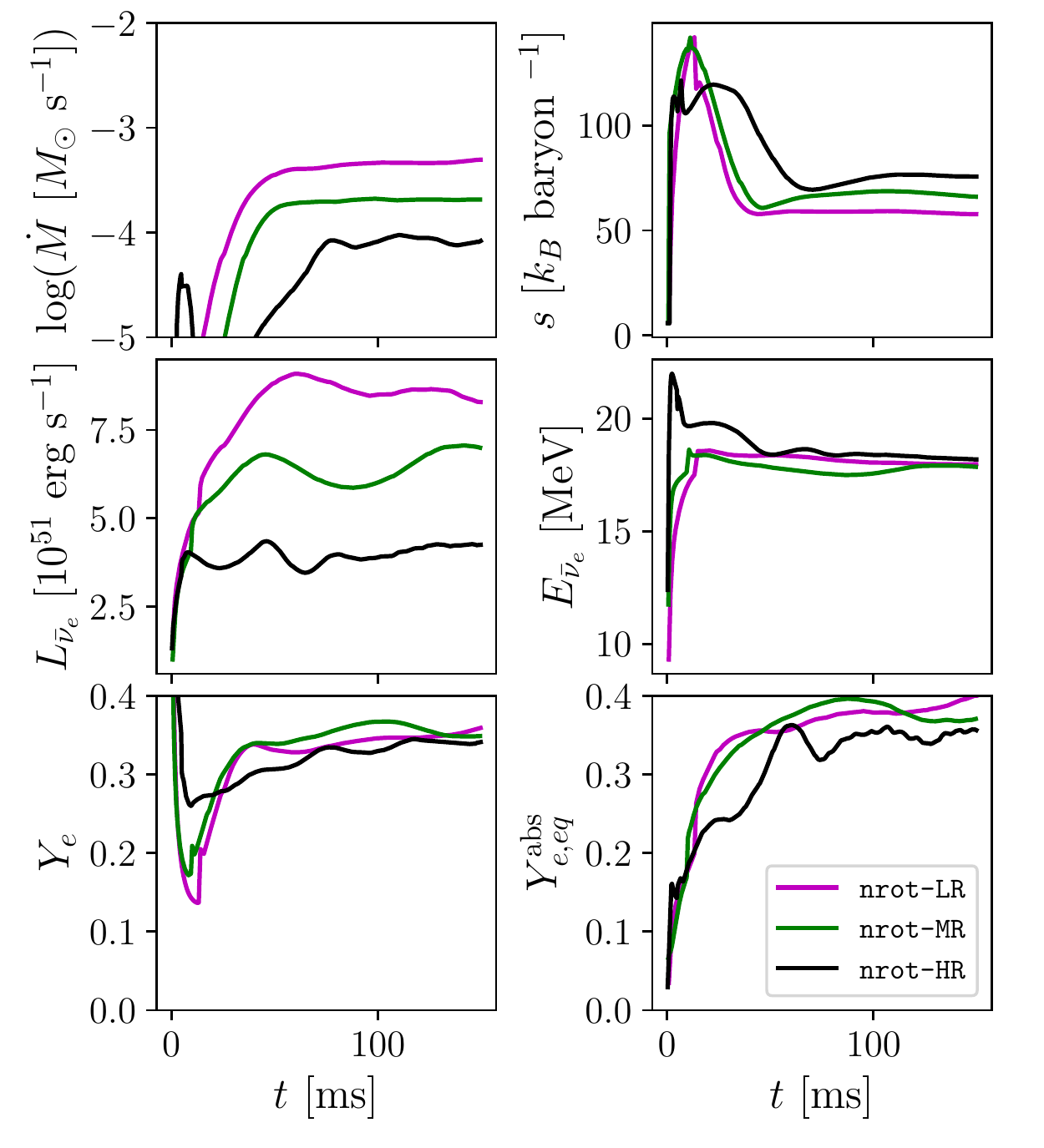}
    \caption{Time evolution of angle-averaged wind properties (as measured through a $r = 60$ km spherical surface) for the non-rotating models of low resolution \texttt{nrot-LR} ({\it purple}), medium resolution \texttt{nrot-MR} ({\it green}), and high resolution \texttt{rot-HR} ({\it black}).  Asymptotic wind properties such as the mass-loss rate and specific entropy are not converged with resolution because the neutrino decoupling region which sets the neutrino radiation field is not resolved (Fig.~\ref{fig:scale_height}).  However, when the wind properties are scaled following the \citet{qian_nucleosynthesis_1996} analytic formulae to results from 1D time-independent models \citet{Thompson+01}, or to each another, based on their respective neutrino luminosities/energies/neutrinosphere radii, they come into better agreement (Table \ref{tab:conv_props}).
    }
    \label{fig:conv_evol}
\end{figure}

\begin{figure*}
    \includegraphics*[width=.95\textwidth]{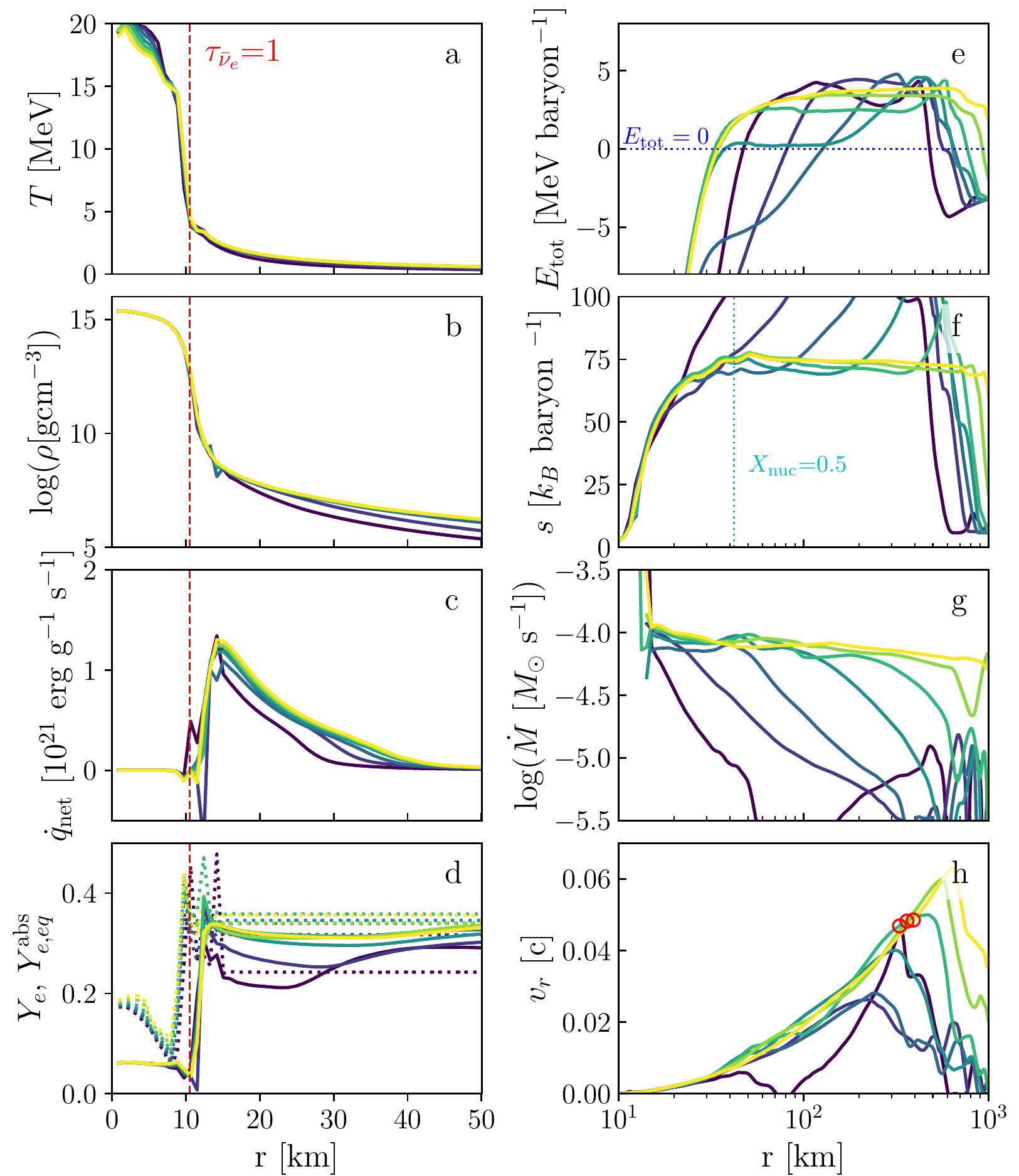}
    \caption{Radial profiles of angle-averaged quantities for model \texttt{nrot-HR} at several times (from dark to light): $t =$30, 49, 68, 87, 105, 124, and 142 ms. Plotted are (a) temperature, $T$; (b) mass density, $\rho$; (c) net specific neutrino heating rate, $\dot q_{\rm net}$; (d) electron fraction, $Y_e$ ({\it solid}), and limited equilibrium electron fraction $Y^{\rm abs}_{e, eq}$ ({\it dotted}); (e) total specific energy, $E_{\rm tot}$; (f) specific entropy, $s$; (g) mass outflow rate, $\dot M$; (h) radial velocity, $v^{r}$.
    The red vertical dashed line denotes the location of the $\bar \nu_e$ neutrinosphere. The $E_{\rm tot}=0$ and $\alpha$-formation surface ($X_{\rm nuc}=0.5$) are indicated by blue and teal dotted lines in panels (e) and (f), respectively. Red circles denote the sonic surface at which the radial velocity $v^{r}$ equals the sound speed.}
    \label{fig:rad_profs}
\end{figure*}

Figures \ref{fig:sph_snaps}--\ref{fig:rad_profs} illustrate our results for the fiducial non-rotating model \texttt{nrot-HR}. Figure~\ref{fig:sph_snaps} shows snapshots through the meridional ($y = 0$) plane of various quantities near the end of the simulation at $t = 152$ ms, once the wind has achieved an approximate steady-state. Figure~\ref{fig:conv_evol} shows the time evolution of the outflowing wind properties, angle-averaged across a spherical surface of radius 60 km. Finally, Fig.~\ref{fig:rad_profs} shows angle-averaged radial profiles of the density $\rho$, temperature $T$, radial velocity $v^{r}$, net specific heating rate $\dot{q}_{\rm net}$, mass-loss rate $\dot{M}$, specific entropy $s$, and electron fraction $Y_e$ at different snapshots in time, starting from $t \approx 30$ ms and going through to the end of the simulation at $t \approx 142$ ms. The net specific heating rate is given by
\begin{eqnarray}
  \dot{q}_{\rm net} = \dot{q}^{+} - \dot{q}^{-} 
  = \sum_{\nu_e, \bar{\nu}_e} \kappa_{\nu_i} (n_{\nu_i}/\rho) E_{\nu_i}
  - \sum_{\nu_i} \dot{q}_{\nu_i}^\mathrm{eff}, \label{eq:qdot_net}
\end{eqnarray}
where $\kappa_{\nu_i}$, $n_{\nu_i}$, and $E_{\nu_i}$ denote the neutrino opacities, number densities, and mean energies, respectively, and $\dot{q}_{\nu_i}^\mathrm{eff}$ are the total specific cooling rates for each neutrino species $\nu_i = \{\nu_e, \bar{\nu}_e, \nu_X\}$. Here, $\nu_X$ collectively labels the heavy-lepton neutrinos and antineutrinos.

After a transient phase, the density profile close to the PNS surface settles into an exponentially declining profile of a hydrostatic atmosphere, which then transitions at larger radii to a more gradual power-law decline characteristic of a wind (Fig. \ref{fig:rad_profs}).  The radii of the $\nu_e$ and $\bar{\nu}_e$ neutrinospheres are denoted by red and blue dots in Fig.~\ref{fig:rad_profs}, respectively.

The radial profiles of the matter density, temperature, and net specific neutrino heating rate ($\dot{q}_{\rm net} = \dot{q}^{+} - \dot{q}^{-}$; heating minus cooling) settle into an approximate steady state by $t \approx 68$ ms (Fig.~\ref{fig:rad_profs}\,a--c). By this time, the neutrino luminosities and energies also stabilize (Fig.~\ref{fig:conv_evol}, middle row), with $L_{\nu_e} \approx 2.5\E{51}\rm~erg~s^{-1}$, $L_{\bar{\nu}_e} \approx 4.2\E{51}\rm~erg~s^{-1}$, $E_{\nu_e} \approx 13$ MeV, and $E_{\bar{\nu}_e} \approx 18$ MeV (Table \ref{tab:models}). Over this same period, the radius of the neutrinosphere (taken to be that of $\bar{\nu}_e$) grows from $\approx 10$ km to 10.5 km.  These PNS luminosities and radii correspond to those achieved on timescales of a few seconds after a successful core collapse supernovae (e.g., \citealt{Pons+99,scheck_multidimensional_2006,Roberts+12a}), the same epoch over which the bulk of the total PNS mass-loss occurs.

The black lines in Fig.~\ref{fig:ang_profs} show the time-averaged wind properties through two different spherical surfaces ($r = 60, 120$ km) as a function of polar angle $\theta$ with respect to the $z$-axis.  With the exception of the isotropic neutrino luminosity right along the pole ($\theta = 0$), most of the wind quantities are roughly spherically symmetric and show no major effects of the grid boundaries, with $\dot{M}$ and $v^{r}$ varying by factors $\lesssim 2$ across all $\theta$ and the other quantities varying by $\lesssim 10\%.$  

Close to the PNS surface, heating from neutrino captures balance cooling from pair captures (Eq.~\ref{eq:ntop}). Moving above the surface, the temperature drops (Fig.~\ref{fig:rad_profs}\,a) and neutrino cooling from pair captures on nuclei ($\dot{q}^{-} \propto T^6$) plummets, while the heating rate $\dot{q}^{+}$ remains roughly constant with radius.  Consequently, a ``gain layer" of net neutrino heating $\dot{q}_{\rm net} = \dot{q}^{+} - \dot{q}^{-} > 0$ forms at radii $r\sim 10-50$ km (Fig.~\ref{fig:rad_profs}\,c).  This heating causes the entropy of the outflowing material to rise, $s= \int dq/T$, in the gain layer and then plateau at larger radii to $s\simeq 74$ (Fig.~\ref{fig:rad_profs}\,f).

The $\alpha$-particle formation surface lies at $r \approx 45$ km, where the temperature has dropped to $\approx 0.5-0.7$ MeV. Heat released during $\alpha$ recombination reactions introduces an increase in the wind entropy, which can be seen as the bright ring in the top right panel of Fig.~\ref{fig:sph_snaps}, or the fluctuations at $r=30-60$ km in the entropy radial profiles of Fig.~\ref{fig:rad_profs}\,f.

Radiation pressure dominates over gas pressure in the high-entropy outflow near and above the gain layer, and the resulting radial pressure gradient causes material to accelerate outwards.  The radial velocity, plotted in Fig.~\ref{fig:rad_profs}\,h, increases with radius.  Material becomes unbound from the PNS around $r \approx\!30-40$\,km, as indicated by a positive total specific energy $E_{\rm tot} = -h u_t -1$ (Figs.~\ref{fig:sph_snaps} and \ref{fig:rad_profs}\,e).

By $t \approx 105$ ms, a sonic surface, at which the radial velocity equals the sound speed, has been established around $r = 300$\,km (denoted by red circles in Fig.~\ref{fig:rad_profs}\,h). By the end of the simulation the sonic surface is approaching $\approx\!400$\,km (see also Fig.~\ref{fig:sph_snaps}) and the wind has attained a velocity $v\gtrsim 0.06c$.  The sonic radius agrees with those found by \citet{Thompson+01} for similar wind parameters.  Though the wind is still accelerating, its total energy $E_{\rm tot}$ has plateaued to a value $\approx 3$ MeV per baryon, which will translate (once the enthalpy is converted into bulk kinetic energy) into an asymptotic speed $v_\infty \equiv \sqrt{2E_{\rm tot}} \approx 0.09$ c. By the final snapshot, the mass loss rate approaches the radially constant profile expected of a steady-state wind (Fig.~\ref{fig:rad_profs}\,g), reaching an asymptotic value $\dot{M} \approx 3.2\times 10^{-4} M_{\odot}$ s$^{-1}$.

Absorption of neutrinos by the wind material also causes the electron fraction $Y_e$ to rise with radius from its low value near the neutrinosphere (Fig.~\ref{fig:conv_evol}, bottom row). Neglecting relativistic effects, the electron fraction in an outflowing fluid element evolves according to the reactions in Eq.~\eqref{eq:ntop},
\be
\frac{dY_e}{dt} = \left(\lambda_{e^+}+\lambda_{\nu_e} \right)\left(1-Y_e \right) -\left(\lambda_{e^-}+\lambda_{\bar \nu_e} \right)Y_e,
\label{eq:dyedt}
\ee
where $\lambda_{e^+}$ and $\lambda_{\nu_e}$ are the positron and electron neutrino capture rates on neutrons, and $\lambda_{e^-}$ and $\lambda_{\bar \nu_e}$ are the electron and electron anti-neutrino capture rates on protons, respectively.  The equilibrium electron fraction at any location can be defined as the value $Y_e = Y_e^{\rm eq}$ for which $dY_e/dt = 0$.  This equilibrium is approached on the characteristic timescale,
\be
\tau_\beta = \frac{1}{\left(\lambda_{e^+}+\lambda_{\nu_e} \right)\left(1-Y_e \right) +\left(\lambda_{e^-}+\lambda_{\bar \nu_e} \right)Y_e}.
\ee
Near the PNS surface, where the radial velocity is low, $\tau_\beta$ is much shorter than the expansion time of the outflow ($\tau_{\rm exp} \sim \rho/ \dot \rho$) and hence $Y_e \simeq Y_e^{\rm eq}$ is well-satisfied.  However, as the wind accelerates at larger radii, $\tau_{\rm exp}$ decreases, until eventually $\tau_{\beta} \gtrsim \tau_{\rm exp}$, causing $Y_e$ to freeze out.

In PNS winds, the temperatures are sufficiently low by the radii at which $Y_e$ freeze-out occurs that the pair capture reactions ($\lambda_{e^+}$, $\lambda_{e^-}$) are negligible compared to the neutrino absorption reactions ($\lambda_{\nu_e}$, $\lambda_{\bar{\nu}_e}$).  Thus, around the point of freeze-out, a limited equilibrium has been achieved, in which $\nu_e$ absorption reactions on protons balance $\bar{\nu}_e$ absorption reactions on neutrons.\footnote{The fact that $Y_e \simeq Y_{e,eq}^{\rm abs}$ in PNS winds can also be understood from an energetic argument (e.g., \citealt{Metzger+08}): (1) the wind is unbound from the gravitational potential well of the PNS by neutrino heating; (2) because the gravitational binding energy per nucleon $\sim 200$ MeV greatly exceeds the mean energy of the neutrinos absorbed by the wind material $\lesssim 20$ MeV, each nucleon must absorb several neutrinos on average to become unbound; (3) from these multiple absorptions per nucleon, the wind necessarily ``forgets'' about the initial ratio of protons to neutrons on the PNS surface in favor of $Y_e^{\rm eq}$.}   The electron fraction corresponding to this limited equilibrium, $Y_{e,eq}^{\rm abs}$, depends exclusively on the $\nu_e/\bar{\nu}_e$ neutrino radiation fields (\citealt{Qian+93}, \citealt{Qian&Woosley96}), viz.~
\begin{eqnarray}
Y_{e,eq}^{\rm abs} &\simeq& \frac{\lambda_{\nu_e}}{\lambda_{\nu_e}+\lambda_{\bar{\nu}_e}} \nonumber \\
&\approx& \left(1 + \frac{L_{\bar{\nu}_e}}{L_{\nu_e}}\frac{\langle \epsilon_{\bar{\nu}_e}\rangle - 2 \Delta + 1.2\Delta^{2}/\langle \epsilon_{\bar{\nu}_e}\rangle}{\langle \epsilon_{\nu_e}\rangle - 2 \Delta + 1.2\Delta^{2}/\langle \epsilon_{\nu_e}\rangle}\right)^{-1},
\label{eq:Yeeqabs}
\end{eqnarray}
where $\Delta \equiv m_n - m_p \simeq 1.293$ MeV is the proton-neutron mass-difference and $\langle \epsilon_{\nu} \rangle = \langle E_{\nu}^{2} \rangle/\langle E_{\nu} \rangle$ is the corresponding mean neutrino energy.

Figure~\ref{fig:rad_profs}\,d shows that the non-rotating PNS wind achieves $Y_e < Y_{e,eq}^{\rm abs}$ near the PNS surface at early times, but that $Y_e$ approaches $Y_{e,eq}^{\rm abs}$ at large radii and late times.  At larger radii $r \gtrsim 15$ km, both $Y_{e,eq}^{\rm abs}$ and $Y_e$ are roughly constant with radius, as expected since the neutrino radiation field is fixed well above the neutrinosphere and weak interactions have frozen out, respectively. 

Figure~\ref{fig:conv_evol} (bottom row) shows that by $t \sim 100$ ms, $Y_{e,eq}^{\rm abs}$ has settled close to a value of $\sim$0.35, with $Y_e$ approaching this value as well, from below.  These $Y_e$ values are lower than predicted by detailed PNS cooling calculations at epochs of comparable neutrino luminosities to those of our solutions (e.g., \citealt{roberts_medium_2012,martinez-pinedo_charged-current_2012,Pascal+22}); this is not surprising because our initial conditions are not based on the self-consistent outcome of a successful supernova and because the decoupling region which determines the neutrino luminosities and energies is not well-resolved (Fig.~\ref{fig:scale_height}).  

The outflow properties are approximately independent of polar angle and radius, as expected for the spherical outflow of a non-rotating PNS (cf.~Fig.~\ref{fig:ang_profs}).  Table \ref{tab:wind_props} summarizes the asymptotic wind properties, including $\dot M$, $s_{\infty}$, $Y_e$, and $v_{\infty}$.  These are usefully compared to time-independent 1D wind solutions available in the literature (e.g., \citealt{Thompson+01}).  The closest model of \citet{Thompson+01} to our non-rotating models is their $M = 1.4 M_{\odot}$ model with $L_{\nu_e} = 3.1\E{51} \rm erg~s^{-1}$, $L_{\bar{\nu}_e} = 4\E{51}~\rm erg~s^{-1}$, for which they obtain: $\dot M \approx 1.63\E{-5} M_{\odot}$ s$^{-1}$, $s_{\infty} \approx 98$, and $v_{\infty} \approx 0.08c$ (their Table 1; hereafter model \texttt{Thompson}).  As summarized in Table \ref{tab:wind_props}, other than the value of $Y_e$ (which is not expected to agree given the different $L_{\nu_e}/L_{\bar \nu_e}/E_{\nu_e}/E_{\bar \nu_e}$ values), the \citet{Thompson+01} wind properties broadly agree with those of our non-rotating PNS wind solutions.  This agreement is further improved if the wind properties are scaled to ours using the analytic formulae of \citet{Qian&Woosley96} given our solutions' respective neutrino luminosities and energies.  

To verify that our wind properties converge, we compare results from three otherwise similar non-rotating wind simulations with different resolutions in Fig~\ref{fig:conv_evol}.  Figure~\ref{fig:scale_height} shows that we do not resolve the neutrino decoupling region in any of the models, as this would require a tenfold increase in spatial resolution.  As such, the predicted properties of the neutrino radiation field $\{L_{\bar{\nu}_e},L_{\nu_e},E_{\bar{\nu}_e},E_{\nu_e}\}$ and the neutrinosphere radii \{$R_{\bar{\nu}_e}, R_{\nu_e}\}$ vary significantly between the models in Fig~\ref{fig:conv_evol}. However, after scaling the steady-state wind quantities $\dot M$ and $s_{\infty}$ to the closest equivalent model of \citet{Thompson+01} and to each other based on their respective neutrino properties following the analytic formulae from \citet{Qian&Woosley96}, the wind quantities come into approximate agreement (within tens of percent; Table~\ref{tab:conv_props}).

The parameter $\eta = s_{\infty}^{3}/(Y_e^{3}\tau_{\rm exp})$ (Eq.~\eqref{eq:eta}) quantifies the potential to form heavy $r$-process elements at large radii in the wind via the $\alpha$-rich freeze-out mechanism (Sec.~\ref{sec:nucleosynthesis}).  Given the asymptotic wind entropy ($s_{\infty} \approx 74$) and expansion time through the seed-formation region $(\tau_{\rm exp} \approx 21.5$ ms) of our non-rotating model, this yields a value $\eta \approx 4.5\times 10^{8}$ (Table~\ref{tab:wind_properties}), well below the threshold value $\eta_{\rm thr} \approx 9\times 10^{9}$ to achieve even a second-peak $r$-process (Eq.~\eqref{eq:eta}) for values of the electron fraction $Y_e \gtrsim \bar{Z}/\bar{A} \sim 0.4$ regime.  The inability of spherically symmetric non-rotating purely-neutrino-driven winds to yield a successful heavy $r$-process is consistent with previous findings (see discussion and references in Sec.~\ref{sec:intro}).  The next subsection addresses whether these conclusions change in the presence of rapid rotation.    

\begin{table*}
  \begin{center}
    \caption{Asymptotic Properties of Rotating and Non-Rotating PNS Winds}
    \label{tab:wind_properties}
    \begin{tabular}{c|c|c|c|c|c|c} 
    \hline
      Model & $s_{\infty}$ & $Y_e$ & $\eta/\eta_{\rm crit}$ (Eq.~\ref{eq:eta}) & $\tau_{\rm exp}$ & $\dot{M}$  & $v_\infty^{(c)}$\\

      - & [$k_B$ baryon$^{-1}$] & - & - & [ms] & [$M_{\odot}$ s$^{-1}$] & $[c]$ \\
      \hline
      \hline
\texttt{nrot-HR} & 73.9 & 0.34 & $4.9\E{-2}$& 21.5 & $7.40\E{-5}$ & $\gtrsim 0.09$\\
\hline 
\texttt{Thompson}$^{\dagger}$ & 98.4 & 0.47 & $3.9\E{-2}$ & 23.79 & $1.63\E{-5}$ & $\approx 0.08$\\
\hline 
\texttt{nrot-HR} scaled$^{\ddagger}$ to \texttt{Thompson} & 85.1 & 0.47 & $1.9\E{-2}$ & 32.5 & $1.80\E{-5}$ & - \\
\hline
\texttt{rot.6-MR} ($0 \le \theta \le 180^{\circ}$) & 28.6 & 0.33 & $2.2\E{-3}$ & 31.1 & $1.19\E{-3}$ & 0.037  \\
\hline
\texttt{rot.6-MR} (Polar$^{a}$) & 56.9 & 0.35 & $1.4\E{-2}$ & 15.2 & $1.37\E{-4}$ & 0.096 \\
\hline
\texttt{rot.6-MR} (Equatorial$^{b}$) & 19.2 & 0.30 & $1.0\E{-3}$ & 49.3 & $6.09\E{-4}$ & 0.019\\

\hline
    \end{tabular} \\ All quantities are averaged over the final third of the simulation run (e.g., $100-150$ ms for \texttt{nrot-HR}).
    $^{a}$Averaged over $\theta \in [0,30^{\circ}]$ and $\theta \in [150^{\circ},180^{\circ}]$; $^{b}$Averaged over $\theta \in [60^{\circ},120^{\circ}]$; $^{c}$Asymptotic wind velocity, calculated from the total energy of the wind at large radii according to $v_{\infty} \equiv \sqrt{2E_{\rm tot}}$; $^{\dagger}$From the $M = 1.4M_{\odot}$, $L_{\bar{\nu}_e} = 4\E{51}$ erg s$^{-1}$ model of \citet{Thompson+01} (Row 6 of their Table 1); $^{\ddagger}$Expressions for entropy, mass-loss rate, and dynamical time obtained by rescaling the results of model \texttt{nrot-HR} to the neutrino luminosities, energies, and neutrinosphere radii of model \texttt{Thompson} following analytic expressions from \citet{Qian&Woosley96}, e.g.
    $\dot M \propto \sum_{\nu=\nu_e,\bar \nu_e} L_\nu^{5/3} E_\nu^{10/3}R_\nu^{5/3}$; $s_{\infty} \propto \sum_{\nu=\nu_e,\bar \nu_e} L_\nu^{-1/6} E_\nu^{-1/3}R_\nu^{-2/3}$; $\tau_{\rm exp} \propto \sum_{\nu=\nu_e,\bar \nu_e} L_\nu^{-1} E_\nu^{-2}R_\nu. $ \\
    
\label{tab:wind_props}
  \end{center}
\end{table*}

\begin{table*}
  \begin{center}
    \caption{Resolution Study of Non-Rotating PNS Wind Properties}
    \label{tab:conv_props}
    \begin{tabular}{c|c|c|c|c} 
    \hline
      Model & $s_{\infty}$ & $s_{\infty}$  (scaled)${\ddagger}$ & $\dot{M}$ & $\dot{M}$ (scaled)${\ddagger}$ \\

      - & [$k_B$ baryon$^{-1}$] & [$k_B$ baryon$^{-1}$] & [$M_{\odot}$ s$^{-1}$] & [$M_{\odot}$ s$^{-1}$] \\
      \hline
      \hline
\texttt{Thompson}$^{\dagger}$ & 98.4 & - & $1.63\E{-5}$ & - \\
\hline 
\texttt{nrot-HR} & 73.9 & 85.1 & $7.40\E{-5}$ & $1.80\E{-5}$  \\
\hline 
\texttt{nrot-MR} & 67.6 & 90.2 & $2.02\E{-4}$& $2.00\E{-5}$\\
\hline 
\texttt{nrot-LR}  & 58.7 & 85.0 & $3.71\E{-4}$ & $1.89\E{-5}$  \\
\hline

\hline
    \end{tabular} \\
    Quantities are averaged over polar angle $\theta \in [0,180^{\circ}]$, and in time from $t=$ 100 to 150 ms.
    
    $^{\dagger}$From the $M = 1.4M_{\odot}$, $L_{\bar{\nu}_e} = 4\E{51}$ erg s$^{-1}$ model of \citet{Thompson+01} (Row 6 of their Table 1).
    
    ${\ddagger}$Entropy and mass-loss rate scaled to the \texttt{Thompson} model in the same way as described in Table \ref{tab:wind_properties}.
 
  \end{center}
\end{table*}

\begin{figure}
    \includegraphics*[width=.5\textwidth]{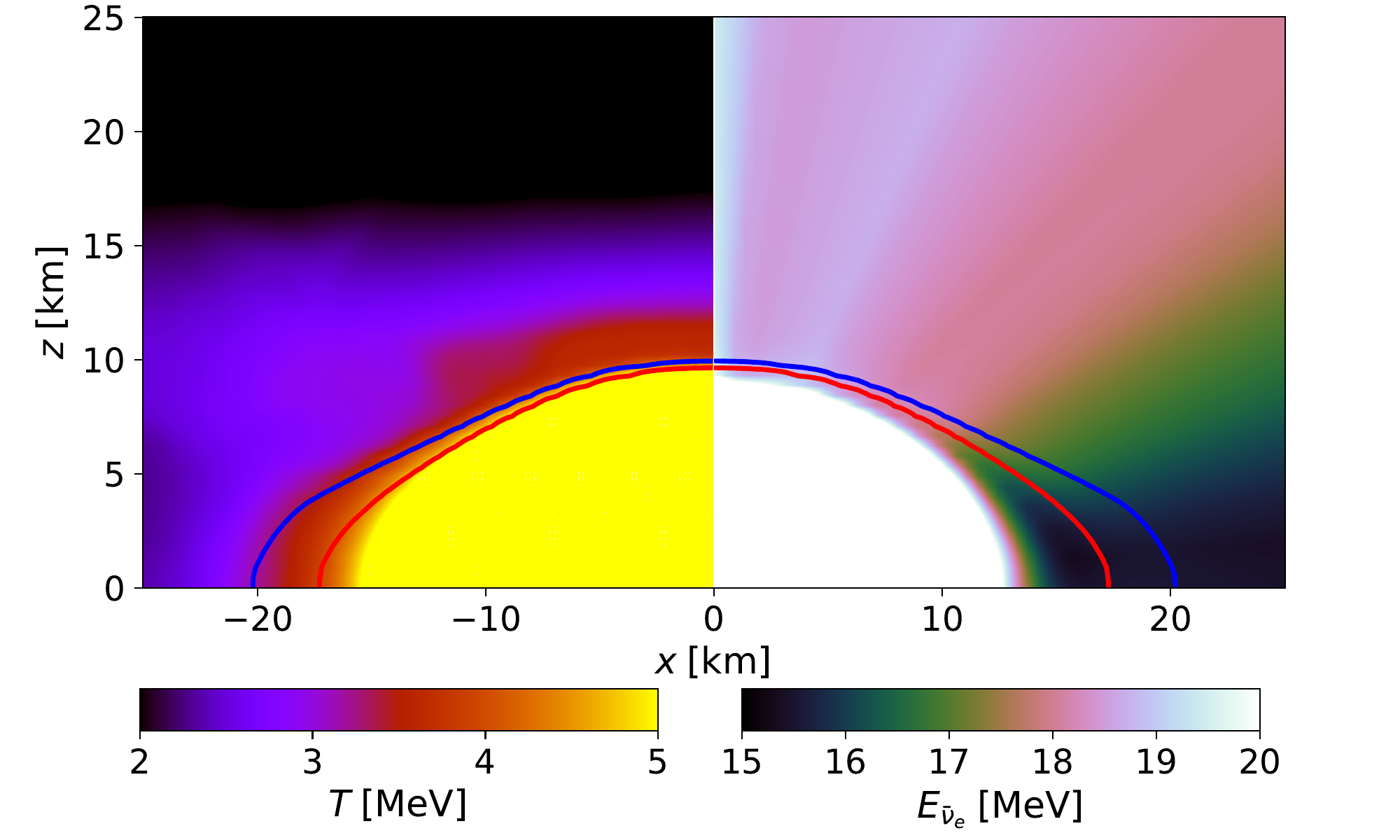}
    \caption{Temperature ({\it left}) and electron anti-neutrino mean energy ({\it right}) close to the PNS surface in the meridional plane for the maximally rotating model \texttt{rot.6-MR}.  The approximate neutrinosphere surfaces (contours $\tau_{\nu_e}$, $\tau_{\bar \nu_e}=1$) for $\nu_e$ and $\bar{\nu}_e$ are shown as blue and red curves, respectively.  A bulge due to rapid rotation pushes out the location of the neutrinosphere near the equator, decreasing the $\tau = 1$ temperatures and reducing the mean neutrino energies in the equatorial regions relative to the polar regions.}
    \label{fig:nuspheres}
\end{figure}

\begin{figure*}
    \includegraphics[width=0.49\textwidth]{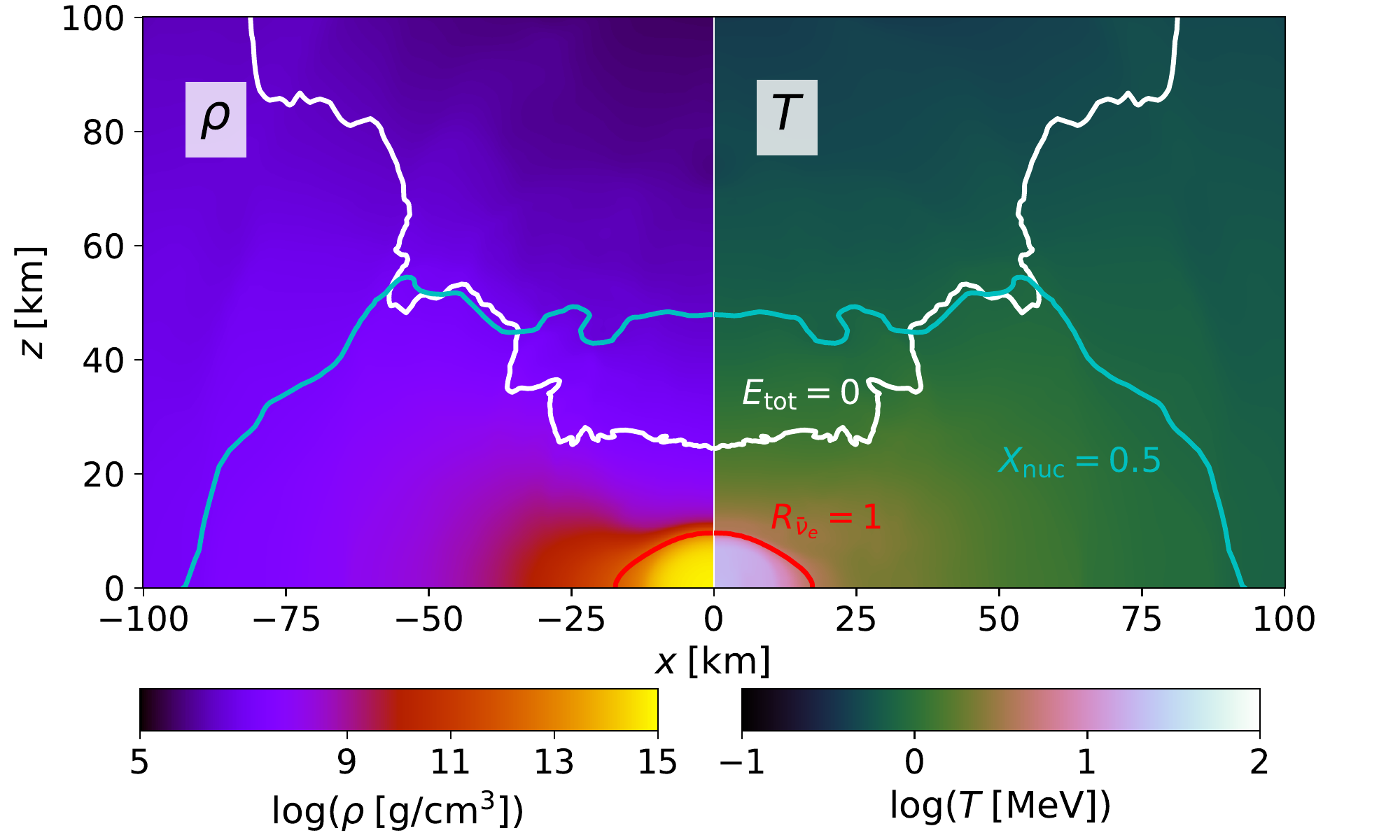}
    \includegraphics[width=0.49\textwidth]{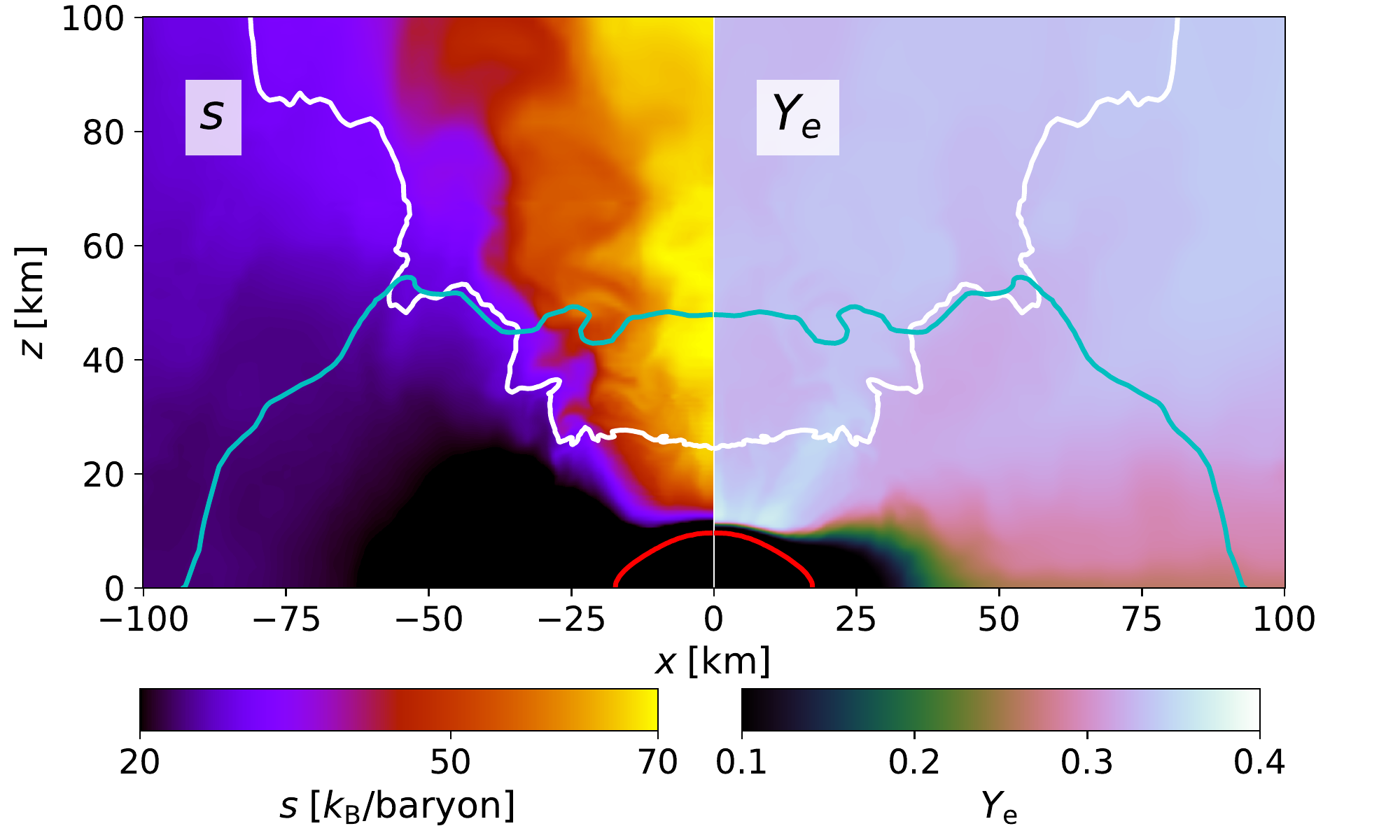}
    \includegraphics[width=0.49\textwidth]{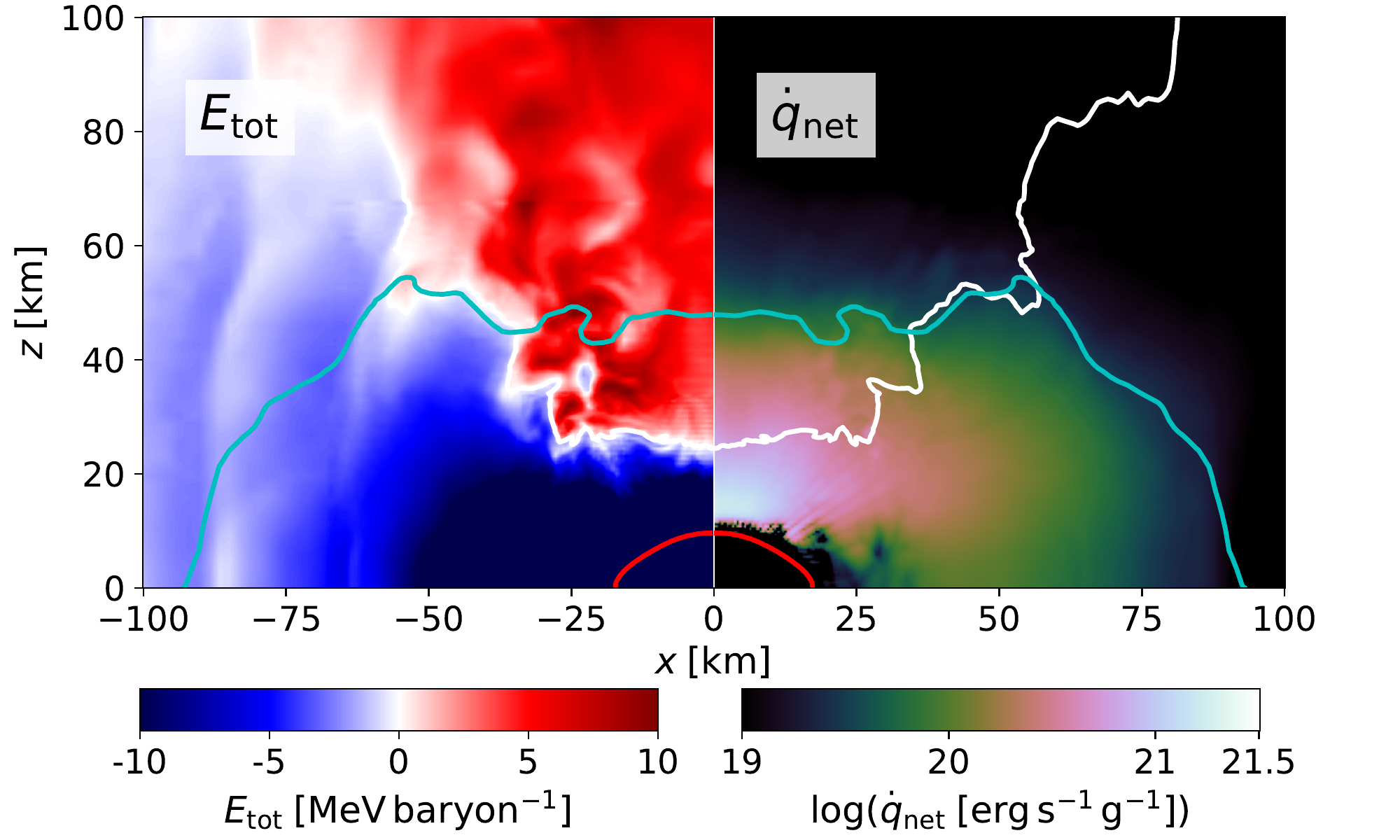}
    \includegraphics[width=0.49\textwidth]{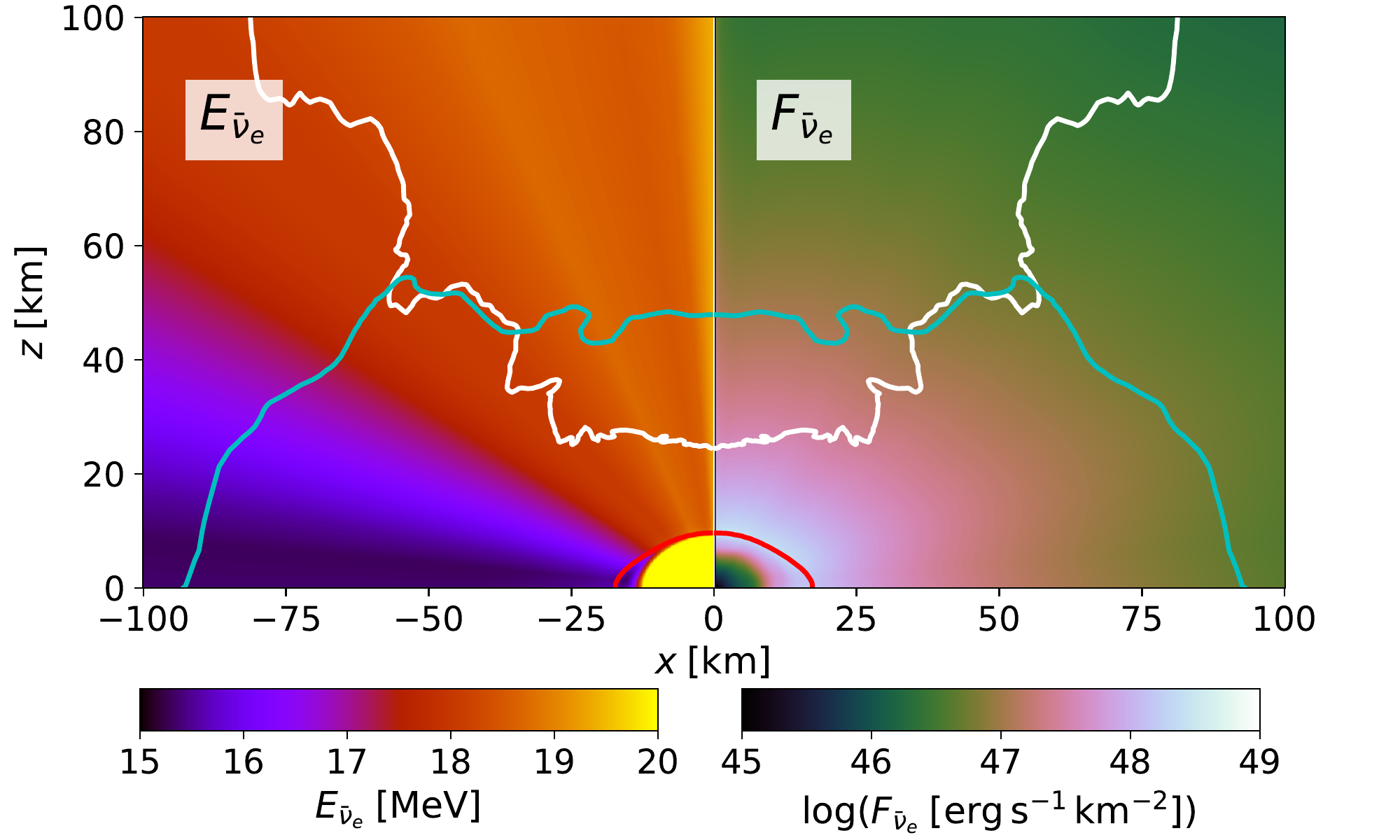}
    \includegraphics[width=0.49\textwidth]{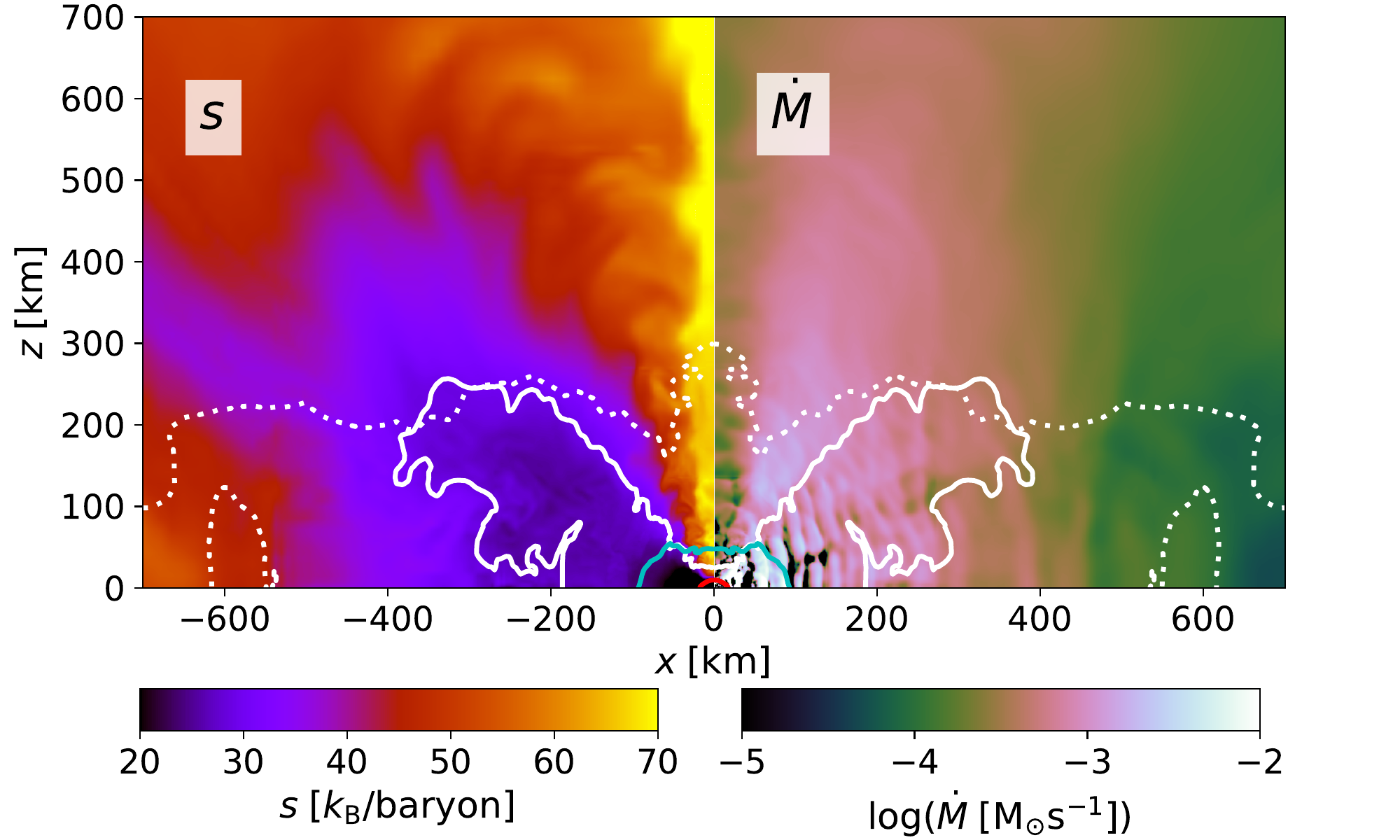}
    \includegraphics[width=0.49\textwidth]{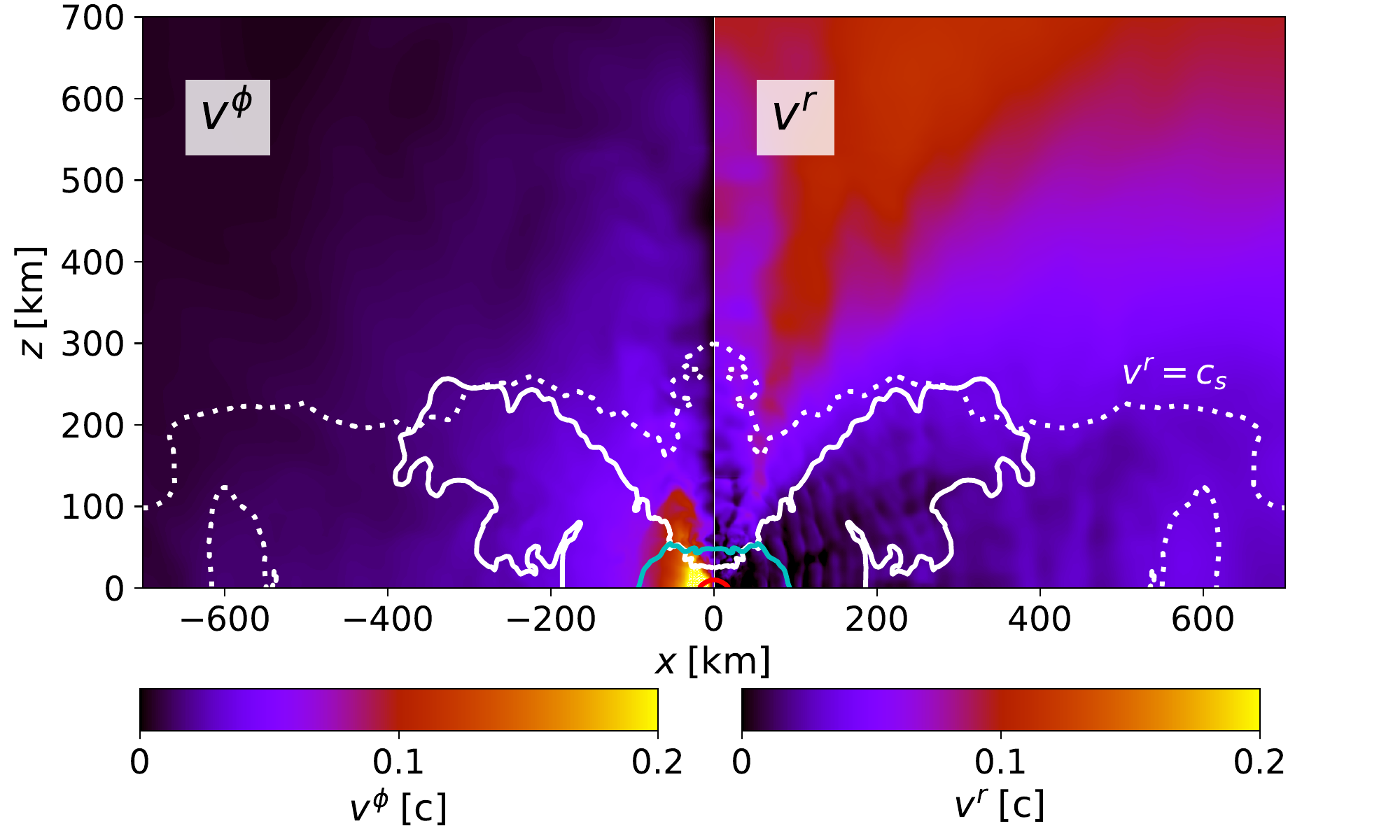}
    \caption{Similar to Fig.~\ref{fig:sph_snaps}, except now showing snapshots at $t = 100$ ms of 2D slices through the rotational axis for model \texttt{rot.6-MR}, and including new quantities, such as azimuthal velocity $v^{\phi}$, electron anti-neutrino flux $F_{\bar \nu_e}$, and mean energy $E_{\bar \nu_e}$.}
\label{fig:rot_snaps}
\end{figure*}

\begin{figure*}
    \centering
    \includegraphics[width=.5\textwidth]{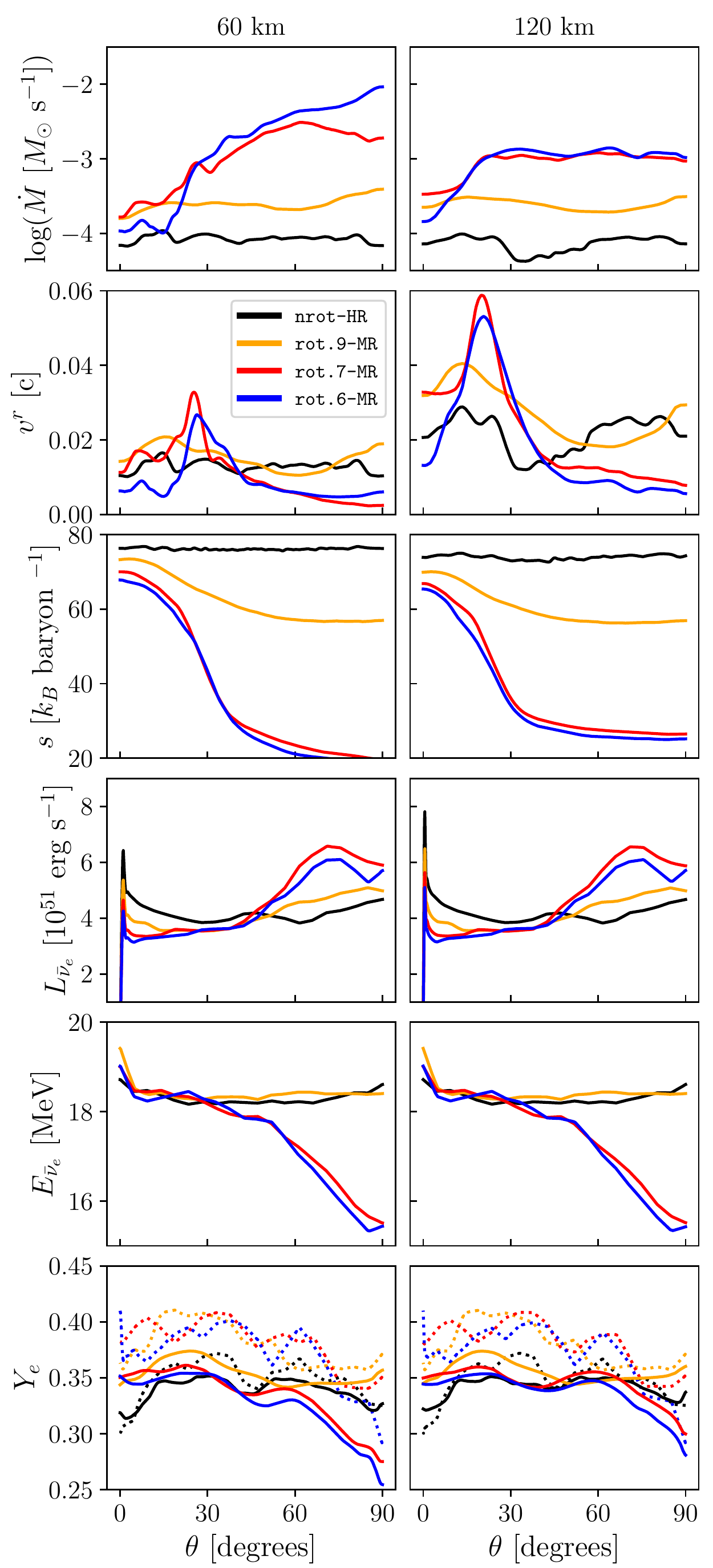}
    \caption{Wind properties as a function of polar angle $\theta$, measured from the $z-$axis (rotation axis in rotating models), through spherical surfaces at $r=60$ km ({\it left}) and $r=120$ km ({\it right}) from the non-rotating model \texttt{nrot-HR} (non-rotating; {\it black}) and rotating models \texttt{rot.9-MR} ({\it orange}), \texttt{rot.7-MR} ({\it red}), and \texttt{rot.6-MR} ({\it blue}), time-averaged over the final third of the simulation run (e.g. $100-150$ ms for $\texttt{nrot-HR}$). From top to bottom, the quantities shown include: isotropic mass-loss rate $\dot M$, radial velocity $v^{r}$, specific entropy $s$, isotropic electron anti-neutrino luminosity $L_{\bar \nu_e}$, mean electron anti-neutrino energy $E_{\bar \nu_e}$, and electron fraction $Y_e$.  Dotted lines in the bottom panel show our estimate of the equilibrium electron fraction $Y_{e,eq}^{\rm abs}$ (Eq.~\eqref{eq:Yeeqabs}).}
    \label{fig:ang_profs}
\end{figure*}

\subsection{Rapidly Rotating PNS Winds}

\begin{figure}
    \includegraphics*[width=.5\textwidth]{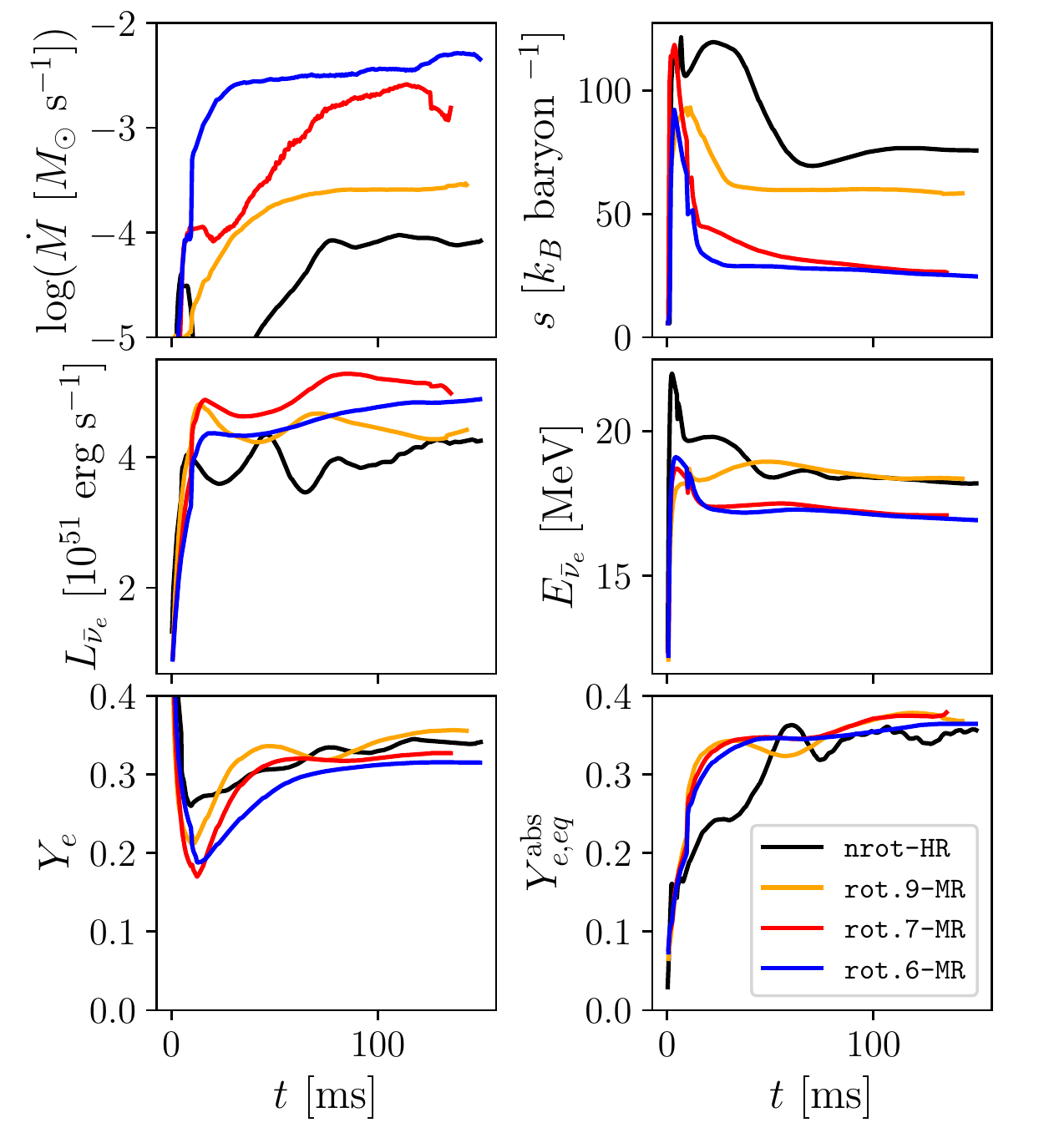}
    \caption{Similar to Fig.~\ref{fig:conv_evol}, except now comparing angle-averaged wind properties in the non-rotating model \texttt{nrot-HR} ({\it black}), and rotating models \texttt{rot.9-MR} ({\it orange}), \texttt{rot.7-MR} ({\it red}), and \texttt{rot.6-MR} ({\it blue}) as measured through a spherical surface of radius $r = 60$ km.
    }
    \label{fig:rot_evol}
\end{figure}

Rotation is expected to have a significant impact on the dynamics of the wind, at least in the equatorial regions of the flow, if the rotational velocity $v^{\phi} \sim R_{\nu}\Omega$ near the neutrinosphere radius $R_{\nu}$ exceeds the sound speed, $c_{\rm s}$, in the gain region.  Gas pressure dominates radiation pressure near the neutrinosphere at the base of the approximately isothermal gain layer ($s \sim 1$; Fig.~\ref{fig:rad_profs}), such that $c_{\rm s} \simeq (kT/m_p)^{1/2}$.  Approximating the single-species neutrino luminosity $L_{\nu} \simeq 4\pi(7/8)R_{\nu}^{2}\sigma T^{4}$ as that of a Fermi-Dirac blackbody, rotation will become dynamically important in the PNS atmosphere for spin periods $P = 2\pi/\Omega$ below a critical value (e.g., \citealt{Thompson+04}),
\be
P_{\rm c} \approx 2\pi \frac{R_{\nu}}{c_s} \approx 3.4\,{\rm ms}\,\left(\frac{R_{\nu}}{12\,{\rm km}}\right)^{5/4}\left(\frac{L_{\nu}}{10^{52}{\rm \,erg\,s^{-1}}}\right)^{-1/8}.
\label{eq:Pc}
\ee

We focus our discussion on the most rapidly-rotating model, \texttt{rot.6-MR}, with axis ratio $R_p/R_e = 0.6$ and spin period of $P = 1.11$ ms $\ll P_{\rm c}$ (Table~\ref{tab:models}).  Figure~\ref{fig:rot_snaps} shows 2D snapshots of various kinematic and thermodynamic quantities in the meridional ($x$-$z$) plane for model \texttt{rot.6-MR} near the end of the simulation at $t=100$ ms. Angular profiles (where $\theta$ is measured from the axis of rotation) of wind quantities through spherical surfaces of radius $r=60$ km and $r = 120$ km are shown in Fig.~\ref{fig:ang_profs}, comparing the results of \texttt{rot.6-MR} to those of the non-rotating model \texttt{nrot-HR} (for which the wind properties are expected and seen to be approximately uniform with angle).

Broadly speaking, our simulations reveal that the rotating wind can be divided into two angular regions with qualitatively distinct properties: (1) a fast polar outflow that develops quickly with properties across this region qualitatively similar to those of a non-rotating PNS; (2) a slower, denser equatorial outflow whose properties differ markedly from the non-rotating case, and which dominates the total mass-loss rate from the star.

After an initial transient phase, the density settles into an approximate steady state with an equatorial bulge of density $\rho \sim 10^{12}\gcc$ extending out to a cylindrical radius $\varrho \approx 30$ km (Fig.~\ref{fig:rot_snaps}), 
compared to the steeper density profile along the polar axis, which falls to $\rho \lesssim 10^{8} \gcc$ by $z\gtrsim 15$ km.  The neutrinosphere surface is likewise oblate in shape, bulging out to $\varrho \approx 17$ km in the equatorial plane compared to $z \approx 10$ km along the polar axis (Fig.~\ref{fig:nuspheres}), the latter being similar to the spherical neutrinosphere radius in the non-rotating model \texttt{nrot-HR}.  The time- and angle-averaged $\nu_e$ and $\bar{\nu}_e$ luminosities of model \texttt{rot.6-MR} are similar to those of \texttt{nrot-HR} (Table \ref{tab:models}).  These differences are small enough that comparing these models allows us to roughly isolate the effects of rotation on the wind properties at a fixed epoch in the PNS cooling evolution (i.e., at approximately fixed neutrino luminosity).

Matter near the PNS surface in the rotational equator has high azimuthal velocity $v^{\phi} \gtrsim 0.2c$ due to the rapid rotation. Moving above the stellar surface, $v^{\phi} \propto 1/\varrho$, consistent with conservation of specific angular momentum in the wind, $\ell = \varrho v^{\phi} = \text{const.}$ (Fig.~\ref{fig:rot_snaps}). 

The isothermal surfaces are also oblate in shape, with a slightly higher neutrinosphere temperature (and hence mean neutrino energy) along the polar axis than in the equator (Fig.~\ref{fig:nuspheres}).  This difference may in part be attributable to the \citet{von_zeipel_radiative_1924} effect, whereby the effective temperature scales with the effective surface gravity $T_{\rm eff} \propto g_{\rm eff}^{1/4}$.  Given that $g_{\rm eff}(R_{\rm eq})/g_{\rm eff}(R_{\rm p}) \sim \left[1-\Omega^{2}/\Omega_{\rm K}^{2}\right] \sim 0.1$ for model \texttt{rot.6-MR}, this would predict $T_{\rm eff}(R_{\rm eq}) \lesssim 0.6T_{\rm p}(R_{\rm eq}),$ close to the ratio of polar and equatorial mean neutrino energies (Fig.~\ref{fig:ang_profs}).  

The gain region around where $\dot{q}_{\rm net}$ peaks is also more radially extended in the equatorial region, but the peak heating rate $\dot{q}_{\rm net}$ is noticeably higher in the polar region, where it again resembles that seen of the non-rotating case.  This enhancement of the polar heating rate results from the greater neutrino flux in this region and the higher mean neutrino energy (Fig.~\ref{fig:rot_snaps}).  The neutrino heating rate drops off abruptly outside of the $\alpha$-particle formation surface (see below), because the neutrino absorption cross section of $\alpha$-particles is much smaller than that of free nucleons. 

The isotropic mass-loss rate $\dot M \sim 10^{-4} M_{\odot}$ s$^{-1}$ along the polar direction in model \texttt{rot.6-MR}, is similar to that of the non-rotating model \texttt{nrot-HR} (Fig.~\ref{fig:ang_profs}).  By contrast, the values of $\dot{M}$ in the equatorial plane are larger than those in the non-rotating case by $1-2$ orders of magnitude, $\sim 10^{-3}-10^{-2} M_{\odot}$ s$^{-1}$.  Rotational enhancement of the mass-loss rate is a well known-effect in thermally-driven winds (e.g., \citealt{Lamers&Cassinelli99}, and references therein).  Rotation has the effect of expanding the density scale-height of the atmosphere $H \approx c_{\rm s}^{2}/g_{\rm eff}$, where $g_{\rm eff} = g - a_c$, and $g$ and $a_c$ are the gravitational and centripetal acceleration experienced by material in the equator, respectively.  A larger scale height exponentially increases the mass in the gain region (since $\rho \propto e^{-H/r}$), thus boosting $\dot{M}$ at the equator relative to the pole, despite the lower specific neutrino heating rate in the equatorial regions.  Latitudinal mixing of the wind material occurs moving outwards with radius; however, an order of magnitude pole-to-equator difference in $\dot{M}$ is preserved to large radii $\gtrsim 120$ km (Fig.~\ref{fig:ang_profs}), outside the sonic surface where further mixing is unlikely to occur.

Another consequence of the lower specific heating rate in the equatorial plane is a suppression of the wind entropy with increasing polar angle.  The entropy along the polar direction, $s \approx 70$, is similar to that of the non-rotating PNS wind solution, compared to $s \approx 20$ in the equatorial plane (Fig.~\ref{fig:ang_profs}). 

Since material in the rotational equator starts out more weakly bound to the star and receives less heating, the radial velocity $v^{r}$ is lower there ($\lesssim 0.05c$) and matter is slower to become unbound.  In the polar region, where the net neutrino heating is maximal, matter accelerates to supersonic velocities $v \approx 0.1 c$ within a few hundred km (Fig.~\ref{fig:rot_snaps}), 
significantly closer to the PNS than in the non-rotating case (Fig.~\ref{fig:sph_snaps}).  This higher polar acceleration may result from ``focusing'' of the polar flow by the denser equatorial outflow (somewhat akin to the `de Laval nozzle' effect; \citealt{Blandford&Rees74}), which causes the areal function of the polar flow-lines to decrease with radius differently than the $\propto 1/r^{2}$ spherical outflow case.  

Matter attains $E_{\rm tot} > 0$ and becomes unbound from the PNS along the polar directions by radii $z \approx 20$ km, while in the equatorial regions this is only achieved outside the $\alpha$-formation surface at radii $\gtrsim$ 90 km (Fig.~\ref{fig:rot_snaps}). 
The significant heating due to $\alpha$-particle formation ($\approx 7$ MeV per nucleon) helps unbind still-marginally bound material, similar to as found in simulations of viscously spreading accretion disks in neutron star mergers (e.g., \citealt{Metzger&Fernandez2013,Siegel&Metzger18}).  Given the low entropy of the outflow, just outside this surface, the $\alpha$-particles rapidly assemble into seed nuclei, releasing further energy.  

The asymptotic value of the wind electron fraction $Y_e$ along the polar directions is $\approx 0.35$, similar or moderately lower than that achieved in the non-rotating model (Fig.~\ref{fig:ang_profs}).  However, outflows from the equator regions are significantly more neutron-rich, with $0.25 \lesssim Y_e \lesssim 0.3$.  In both the polar and equatorial outflow regions, the wind composition still approaches equilibrium with the neutrino radiation field, as evidenced by $Y_e \approx Y_{e,eq}^{\rm abs}$.  The lower value of $Y_{e,eq}^{\rm abs}$ (Eq.~\eqref{eq:Yeeqabs}) and hence $Y_e$ in the equatorial outflow results from the suppression of $L_{\nu_e}$ relative to $L_{\bar{\nu}_e}$ in this region, due to greater $\nu_e$ optical depth through the neutron-rich equatorial bulge.  The large contrast between the temperatures at the $\nu_e$ and $\bar{\nu}_e$ neutrinospheres in the equatorial plane (Fig.~\ref{fig:nuspheres}) gives rise to the distinct average neutrino energies of $\nu_e$ and $\bar{\nu}_e$ along these directions.

\begin{figure}
    \includegraphics[width=0.49\textwidth]{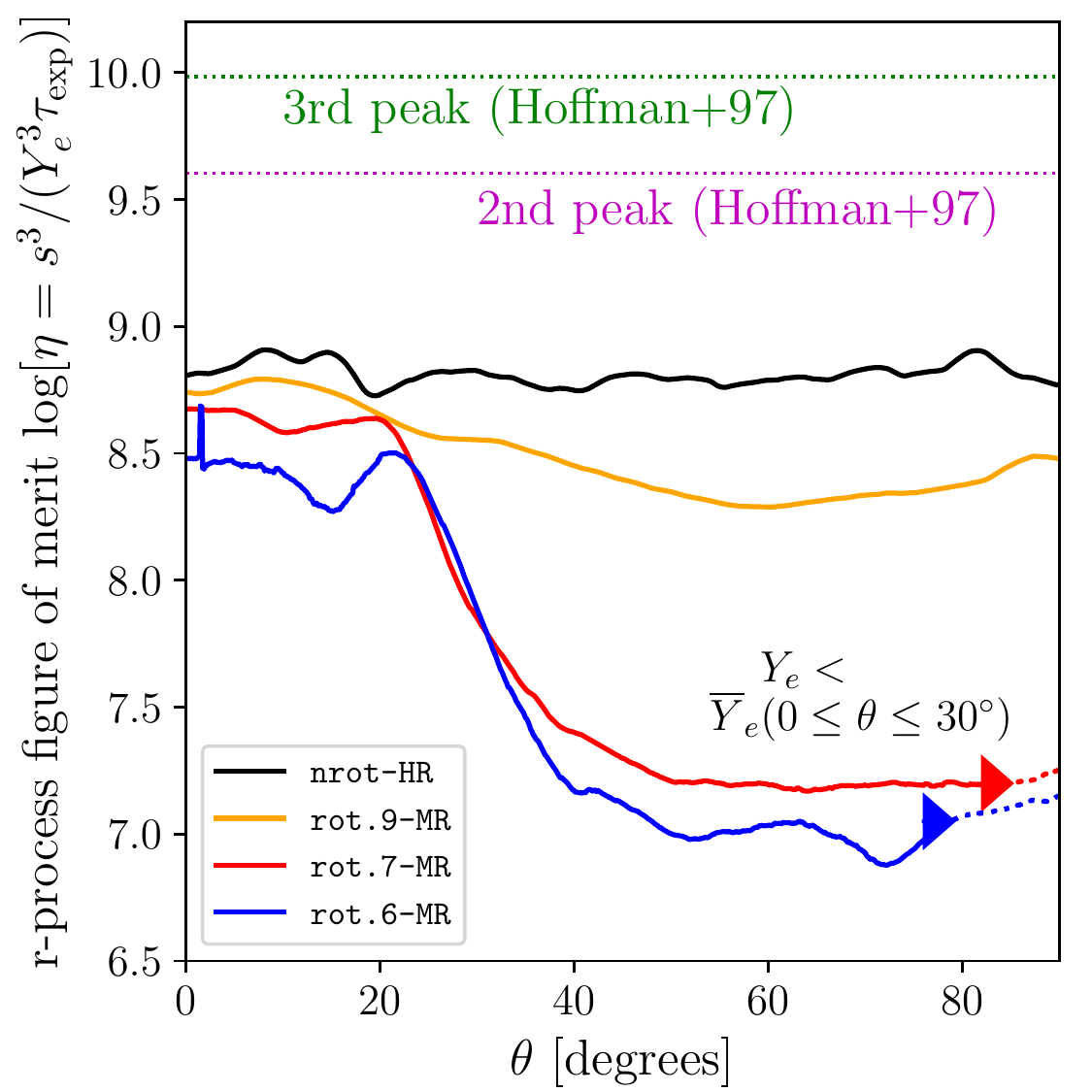}
    \caption{R-process figure-of-merit parameter $\eta \equiv s_{\infty}^{3}/(Y_e^{3}\tau_{\rm exp})$ (Eq.~\eqref{eq:eta}) as a function of outflow polar angle $\theta$, where the relevant quantities are time-averaged and measured through the $T=0.5$ MeV surface, shown separately for the models \texttt{nrot-HR}, \texttt{rot.9-MR}, \texttt{rot.7-MR}, \texttt{rot.6-MR} as marked. For comparison we show the threshold value $\eta_{\rm thr}$ (Eq.~\eqref{eq:eta}) required for neutron captures to reach the 2nd ({\it purple dotted line}) and 3rd ({\it green dotted line}) $r$-process peaks \citep{Hoffman+97}.  A dotted line-style to the right of the solid triangles in the rapidly rotating models (\texttt{rot.6-MR}, \texttt{rot.7-MR}) denotes the angles over which $Y_e$ is 10\% lower than its average value along the polar angles $0^\circ < \theta < 30^\circ$ (the latter is roughly similar to that obtained from a non-rotating wind for the same neutrino emission properties); the outflow from such regions may be capable of a successful $r$-process even absent an $\alpha$-rich freeze-out (i.e., even if $\eta \ll \eta_{\rm thr}$).}
\label{fig:eta_prof}
\end{figure}

The strong angular dependence of several wind quantities also impacts the angular averaged wind properties as illustrated in Fig.~\ref{fig:rot_evol} (see also Tab.~\ref{tab:wind_properties}). Notably, the overall mass-loss rate is enhanced by 1-2 orders of magnitude for the most rapidly rotating model relative to the non-rotating model, while the overall entropy of the wind decreases to less than half of the value of the non-rotating model. As expected from the angular trends in Fig.~\ref{fig:ang_profs}, Fig.~\ref{fig:rot_evol} also shows somewhat larger electron anti-neutrino luminosities, smaller electron anti-neutrino mean energies, and a smaller overall $Y_e$.

As a result of the strong angular dependence of $\{Y_e$, $v^{r}$, $s$, $T\}$ in the rotating models, the $r$-process figure of merit $\eta$ also varies as a function of polar angle $\theta$.  Figure~\ref{fig:eta_prof} compares the angular profile of $\eta$ from the rotating star simulations to the non-rotating model. In the polar region, the rotating and non-rotating models achieve a similar value $\eta \approx 3-6\E{8}$. However, in the equatorial region, $\eta$ is significantly suppressed for the rotating models, with $\eta$ being $\approx$ 2 orders of magnitude lower for $\texttt{rot.7-MR}$ and $\texttt{rot.6-MR}$. Taken together, the value of $\eta$ lies well below the minimum threshold for 2nd or 3rd peak $r$-process element production for all outflow angles and all models.  This disfavors rapidly spinning PNS as $r$-process sources via the $\alpha$-rich freeze-out mechanism. 

At face value, rotation appears to be detrimental to the $r$-process in PNS winds.  However, this does not account for the effect of a lower $Y_e$ alone, absent an $\alpha$-rich freeze-out.  The value of $Y_e$ in our most rapidly rotating models is $\approx 10\%$ smaller in the equatorial direction than along the pole (we denote the $Y_e$-suppressed region with a dashed linestyle in Fig.~\ref{fig:eta_prof}) or in the non-rotating wind model with otherwise similar neutrino luminosities and energies.  Thus, for example, if the ``true'' wind electron fraction at a given point in the cooling evolution of a non-rotating PNS were $Y_{e,0} \approx 0.45-0.5,$ (e.g., \citealt{Roberts+12a}; their Fig.~5), rotation could act to reduce $Y_e$ to $\approx 0.9Y_{e,0} \approx 0.4-0.45$, sufficient to produce neutron-rich light element primary-process (LEPP) nuclei with $38 < Z < 47$ (e.g., \citealt{Arcones&Montes11}), even absent an $\alpha$-rich freeze-out (i.e., even for arbitrarily low values of $\eta$).

\section{Summary and Conclusions}
\label{sec:conclusions}

We have explored the effects of rapid rotation on the properties of neutrino-heated PNS winds by means of three-dimensional GRHD simulations with M0 neutrino transport.  Our conclusions can be summarized as follows.

\begin{itemize}

\item{We calculate a suite of 1.4$M_{\odot}$ PNS models corresponding to different solid-body rotation rates (Tab.~\ref{tab:models}), ranging from the non-rotating case studied in most previous works ($\Omega = 0$) to stars rotating near break-up ($\Omega/\Omega_{\rm K} \simeq 0.94$; $P \simeq 1.11$ ms).  We initialize the axisymmetric PNS structure using the $\texttt{RNS}$ code integrated via a novel procedure with the SFHo tabulated EOS.  Rather than following the self-consistent cooling evolution of the PNS from an initial post-explosion or post-merger state, we initialize the PNS temperature and $Y_e$ radial profiles in $\beta$-equilibrium following \citet{Kaplan+14}.  The chosen temperature normalization generates steady-state neutrino luminosities and energies close to those achieved on a timescale of seconds after a supernova explosion, and over which the bulk of the integrated wind mass-loss will occur.}

\item{Our focus is on studying the wind properties in the gain layer above the PNS surface and out to large radii ($\sim 1000$ km); however, we are not able to fully resolve the neutrinosphere decoupling region (Fig.~\ref{fig:scale_height}).  As a consequence of this, as well as of our idealized initial temperature/$Y_e$ structure, the partitioning between $\nu_e$ and $\bar{\nu}_e$ luminosities and their energies in our simulations do not match those predicted by supernova simulations, nor the resulting wind electron fractions $Y_e \simeq Y_{e,eq}^{\rm abs}$($L_{\nu_e, \bar \nu_e}$, $E_{\nu_e, \bar \nu_e}$).  Specifically, our non-rotating wind solutions achieve values $Y_e \simeq 0.34$ (Fig.~\ref{fig:conv_evol}) significantly lower than those found by detailed PNS cooling calculations, $Y_e \approx 0.45-0.55$ (e.g., \citealt{Roberts&Reddy17,Pascal+22}).  Nevertheless, our simulations can still be used to explore the relative effects of rotation on $Y_{e}$ through a comparison to otherwise equivalent non-rotating models with similar neutrino emission properties.}

\item{After an initial transient phase, all of our models reach approximately steady outflow solutions with positive energies $E_{\rm tot} >0$ and sonic surfaces captured on the computational grid, on timescales $\sim 100$ ms (Fig.~\ref{fig:sph_snaps}, \ref{fig:rad_profs}).  We validate our non-rotating solutions by comparing them to time-independent spherical wind calculations (Tab.~\ref{tab:conv_props}).  Scaling our wind properties ($\dot M$, $s_{\infty}$, $\tau_{\rm exp}$, $v_{\infty}$) based on analytic expressions from \citet{Qian&Woosley96} given the relative neutrino properties ($L_\nu$, $E_\nu$, $R_\nu$), we obtain good agreement (to within $\lesssim 10\%$) with \citet{Thompson+01}.}

\item{Our non-rotating and slower rotating models (\texttt{rot.9-MR}; $\Omega/\Omega_{\rm K} \simeq 0.4$) exhibit approximately spherical outflow properties.  In contrast, the fastest rotating models (\texttt{rot.7-MR}, \texttt{rot.6-MR}; $\Omega/\Omega_{\rm K} \gtrsim 0.75$) generate outflows with distinct properties near the equatorial plane versus higher latitudes closer to the rotational axis (Fig.~\ref{fig:rot_snaps}, Table \ref{tab:wind_properties}).  The outflow properties along the rotational axis ($\theta \lesssim 30^{\circ}$) are qualitatively similar to those of the slowly rotating models in their key properties (e.g., $\dot M$, $Y_e$, $s_{\infty}$, $L_\nu$, $E_\nu$; Fig.~\ref{fig:ang_profs}), as would be expected because centrifugal effects are weak along these directions.  However, outflows from intermediate latitudes accelerate faster to higher speeds compared to a spherical wind (Fig.~\ref{fig:ang_profs}); these features may result from `de Laval'-like focusing of polar streamlines by the denser equatorial outflow. }

\item{The equatorial outflows from rapidly rotating PNS exhibit qualitative differences from the non-rotating case (Fig.~\ref{fig:ang_profs}), as expected because centrifugal forces have a large effect on the hydrostatic structure of the atmosphere for spin periods $P \ll P_{\rm c} \approx 3$ ms (Eq.~\eqref{eq:Pc}).  Relative to slowly rotating models, the equatorial outflows from rapid rotators possess: higher mass-loss rates $\dot{M}$ by over an order of magnitude in the fastest spinning case; slower acceleration and lower asymptotic radial velocities; and lower entropy $s_{\infty}$ by a factor up to $\approx 4$.  These features may be understood as a consequence of the rotation-induced reduction in the effective gravitational mass, when applied to analytic predictions for the $M$-dependence of the wind properties \citep{qian_nucleosynthesis_1996}.

The equatorial outflows of the rapidly rotating solutions are also characterized by lower neutrino energies and a larger contrast between the $\nu_e$ and $\bar{\nu}_e$ neutrinospheres and their respective temperatures/luminosities; these features result from the presence of a dense neutron-rich equatorial bulge/decretion disk near the surface of the star (Fig.~\ref{fig:nuspheres}).  These changes in the $\nu_e/\bar{\nu}_e$ properties reduce the equilibrium electron fraction $Y_{e,eq}^{\rm abs}$ (Eq.~\eqref{eq:Yeeqabs}; and hence $Y_e \lesssim 0.3$) of the equatorial outflows relative to the slowly rotating case.}

\item{Rapid rotation tends to reduce $s_{\infty}$ and to increase $\tau_{\rm exp}$ as a result of the slower expansion speed in the dense equatorial outflow; both effects act to reduce the key parameter $\eta= s_{\infty}^3/(\tau_{\rm exp} Y_e^3)$ (Eq.~\eqref{eq:eta}) by over an order of magnitude in the rotating wind case (Fig.~\ref{fig:eta_prof}).  We conclude that rotation (at least absent a strong magnetic field) does not facilitate a successful 2nd or 3rd $r$-process via the $\alpha$-rich freeze-out mechanism.

On the other hand, outflows near the equatorial plane in our fastest rotating models possess $Y_e$ smaller by $\approx 10-15\%$ compared to the otherwise equivalent non-rotating case.  The winds from very rapidly spinning PNS could therefore generate nucleosynthetic abundance patters which are quantitatively distinct from those of slowly rotating PNS, even neglecting potential rotation-induced changes to the PNS cooling evolution.  For example, if slowly rotating PNS winds achieve $Y_e \gtrsim 0.5-0.55$ (e.g., \citealt{Pascal+22}) and generate mostly iron group elements and $p$-nuclei via the $rp$-process and $\nu p$-process (e.g., \citealt{Frohlich+06,Roberts+10,fischer_protoneutron_2010}), rotating PNS winds could obtain $Y_e \lesssim 0.5$ and would instead synthesize neutron-rich LEPP or light $r$-process nuclei (e.g., \citealt{Qian&Wasserburg07,Arcones&Montes11}).}
\item{The neutrino luminosities achieved by our models $\sim$few $\E{51}$ erg s$^{-1}$ will last for a timescale $\tau_{\rm c} \approx 3$ seconds after a successful supernova explosion (\citealt{Roberts&Reddy17}; their Fig. 3).  Although our simulations are not run this long, our most rapidly spinning PNS solutions therefore predict a total wind-ejecta mass $\sim \dot{M}\tau_{\rm c} \approx 3\times 10^{-3}M_{\odot}$ comprised of more neutron-rich nuclei than would accompany the birth of a slowly rotating PNS of otherwise similar properties (which produce only $\approx 10^{-4}M_{\odot}$ in total wind-ejecta; e.g., \citealt{Thompson+01}).  

Several strains of observational (e.g., \citealt{FaucherGiguere&Kaspi06,Vink&Kuiper06,Perna+08}) and theoretical (e.g., \citealt{ma_angular_2019}) evidence indicate that the birth of neutron stars with rapid spin periods $P \approx 1$ ms are rare in nature among the core collapse population.  However, given their larger wind ejecta mass-yields (by a factor $\gtrsim 10$), even if such rapidly spinning PNS are formed in only $\sim 10\%$ of all core collapse supernovae, their total nucleosynthetic contribution may be competitive with ``ordinary'' supernovae birthing slowly spinning PNS.  Broad-lined supernovae with atypically large ejecta kinetic energies (hinting at an important role of rotation in facilitating the explosion) indeed represent $\sim 10\%$ of core collapse explosions (e.g., \citealt{Perley+20}).  The contributions of rapidly spinning PNS on individual ``pollution events'' observed in the surface abundances of halo stars (e.g., \citealt{Honda+06,Spite+18}) will be further enhanced if stellar cores retain greater angular momentum at core collapse at lower metallicity (e.g., \citealt{yoon_evolution_2005}).

Hot, rapidly spinning PNS-like stars are also generated from the merger of binary neutron stars (e.g., \citealt{Dessart+09}), albeit with higher masses $\gtrsim 2M_{\odot}$ than assumed in our models.  However, the limited lifetimes of most such objects before they lose rotational support and collapse into a black hole, may limit the contribution of their neutrino-driven winds relative to other sources of mass ejection during the merger and its aftermath (though strong magnetic fields may change this picture; e.g., \citealt{siegel_magnetically_2014,Metzger+18,Curtis+21}).  
}

\item{We have treated neutrino transport using an M0 scheme, which neglects the effects lateral transport.  This is likely a good approximation in our case because deviations from spherical symmetry are fairly modest and (compared, e.g. to simulations of the supernova explosion; \citealt{Skinner+16}) we are mainly interested in the properties of the outflows above the neutrino decoupling region.  However, future work should aim to explore the impact of more accurate neutrino transport in the rapidly rotating cases.}

\item{The study presented here lays the groundwork for the future 3D simulation work including additional physical effects.  One of the most important are those arising from strong, ordered magnetic field, which may accompany the birth of rapidly spinning PNS as a result of dynamo processes which tap into the energy available in rotation or convection (e.g., \citealt{Thompson&Duncan93,siegel_magnetorotational_2013,Mosta+14,Raynaud+20}).  Magnetic fields of strength $\gtrsim 10^{14}-10^{15}$ G comparable to those of Galactic magnetars have been shown to have major effects on the PNS wind properties and their efficacy in generating $r$-process elements, both with (e.g., \citealt{Thompson+04,Metzger+07,Winteler+12,Vlasov+14,Vlasov+17}) and without (e.g., \citealt{Thompson03,Thompson&udDoula18}) rapid rotation.}
\end{itemize}

\vspace{0.2cm}

We thank Erik Schnetter and Ben Margalit for discussions and support. This research was enabled in part by support provided by SciNet (www.scinethpc.ca) and Compute Canada (www.computecanada.ca). DD and BDM acknowledge support from the National Science Foundation (grant \#AST-2002577). DMS acknowledges the support of the Natural Sciences and Engineering Research Council of Canada (NSERC), funding reference number RGPIN-2019-04684. Research at Perimeter Institute is supported in part by the Government of Canada through the Department of Innovation, Science and Economic Development Canada and by the Province of Ontario through the Ministry of Colleges and Universities.


\begin{thebibliography}{}
\expandafter\ifx\csname natexlab\endcsname\relax\def\natexlab#1{#1}\fi
\providecommand{\url}[1]{\href{#1}{#1}}
\providecommand{\dodoi}[1]{doi:~\href{http://doi.org/#1}{\nolinkurl{#1}}}
\providecommand{\doeprint}[1]{\href{http://ascl.net/#1}{\nolinkurl{http://ascl.net/#1}}}
\providecommand{\doarXiv}[1]{\href{https://arxiv.org/abs/#1}{\nolinkurl{https://arxiv.org/abs/#1}}}

\bibitem[{{Arcones} \& {Janka}(2011)}]{Arcones+11}
{Arcones}, A., \& {Janka}, H.~T. 2011, \aap, 526, A160,
  \dodoi{10.1051/0004-6361/201015530}

\bibitem[{{Arcones} {et~al.}(2007){Arcones}, {Janka}, \& {Scheck}}]{Arcones+07}
{Arcones}, A., {Janka}, H.~T., \& {Scheck}, L. 2007, \aap, 467, 1227,
  \dodoi{10.1051/0004-6361:20066983}

\bibitem[{{Arcones} \& {Montes}(2011)}]{Arcones&Montes11}
{Arcones}, A., \& {Montes}, F. 2011, \apj, 731, 5,
  \dodoi{10.1088/0004-637X/731/1/5}

\bibitem[{Babiuc-Hamilton {et~al.}(2019)Babiuc-Hamilton, Brandt, Diener, Elley,
  Etienne, Ficarra, Haas, \& Witek}]{babiuc-hamilton_einstein_toolkit_2019}
Babiuc-Hamilton, M., Brandt, S.~R., Diener, P., {et~al.} 2019, The {Einstein}
  {Toolkit} ({The} "{Mayer}" release, {ET}\_2019\_10),  Zenodo,
  \dodoi{10.5281/zenodo.3522086}

\bibitem[{Beloborodov(2003)}]{beloborodov_nuclear_2003}
Beloborodov, A.~M. 2003, \apj, 588, 931, \dodoi{10.1086/374217}

\bibitem[{Beloborodov(2010)}]{beloborodov_collisional_2010}
---. 2010, \mnras, 407, 1033, \dodoi{10.1111/j.1365-2966.2010.16770.x}

\bibitem[{{Bhattacharya} {et~al.}(2021){Bhattacharya}, {Horiuchi}, \&
  {Murase}}]{Bhattacharya+21}
{Bhattacharya}, M., {Horiuchi}, S., \& {Murase}, K. 2021, arXiv e-prints,
  arXiv:2111.05863.
\newblock \doarXiv{2111.05863}

\bibitem[{{Blandford} \& {Rees}(1974)}]{Blandford&Rees74}
{Blandford}, R.~D., \& {Rees}, M.~J. 1974, \mnras, 169, 395,
  \dodoi{10.1093/mnras/169.3.395}

\bibitem[{Bruenn(1985)}]{bruenn_stellar_1985}
Bruenn, S.~W. 1985, \apjs, 58, 771, \dodoi{10.1086/191056}

\bibitem[{Bucciantini {et~al.}(2007)Bucciantini, Quataert, Arons, Metzger, \&
  Thompson}]{bucciantini_magnetar-driven_2007}
Bucciantini, N., Quataert, E., Arons, J., Metzger, B.~D., \& Thompson, T.~A.
  2007, \mnras, 380, 1541, \dodoi{10.1111/j.1365-2966.2007.12164.x}

\bibitem[{{Burrows} \& {Lattimer}(1986)}]{Burrows&Lattimer86}
{Burrows}, A., \& {Lattimer}, J.~M. 1986, \apj, 307, 178,
  \dodoi{10.1086/164405}

\bibitem[{{Burrows} {et~al.}(2020){Burrows}, {Radice}, {Vartanyan}, {Nagakura},
  {Skinner}, \& {Dolence}}]{Burrows+20}
{Burrows}, A., {Radice}, D., {Vartanyan}, D., {et~al.} 2020, \mnras, 491, 2715,
  \dodoi{10.1093/mnras/stz3223}

\bibitem[{{Cardall} \& {Fuller}(1997)}]{Cardall&Fuller97}
{Cardall}, C.~Y., \& {Fuller}, G.~M. 1997, \apjl, 486, L111,
  \dodoi{10.1086/310838}

\bibitem[{Colella \& Woodward(1984)}]{colella_piecewise_1984}
Colella, P., \& Woodward, P.~R. 1984, Journal of Computational Physics, 54,
  174, \dodoi{10.1016/0021-9991(84)90143-8}

\bibitem[{{Curtis} {et~al.}(2021){Curtis}, {M{\"o}sta}, {Wu}, {Radice},
  {Roberts}, {Ricigliano}, \& {Perego}}]{Curtis+21}
{Curtis}, S., {M{\"o}sta}, P., {Wu}, Z., {et~al.} 2021, arXiv e-prints,
  arXiv:2112.00772.
\newblock \doarXiv{2112.00772}

\bibitem[{{Dessart} {et~al.}(2009){Dessart}, {Ott}, {Burrows}, {Rosswog}, \&
  {Livne}}]{Dessart+09}
{Dessart}, L., {Ott}, C.~D., {Burrows}, A., {Rosswog}, S., \& {Livne}, E. 2009,
  \apj, 690, 1681, \dodoi{10.1088/0004-637X/690/2/1681}

\bibitem[{{Duan} \& {Qian}(2004)}]{Duan&Qian04}
{Duan}, H., \& {Qian}, Y. 2004, Phs.~Rev.~D, 69, 123004,
  \dodoi{10.1103/PhysRevD.69.123004}

\bibitem[{{Duncan} {et~al.}(1986){Duncan}, {Shapiro}, \&
  {Wasserman}}]{Duncan+86}
{Duncan}, R.~C., {Shapiro}, S.~L., \& {Wasserman}, I. 1986, \apj, 309, 141,
  \dodoi{10.1086/164587}

\bibitem[{{Faucher-Gigu{\`e}re} \& {Kaspi}(2006)}]{FaucherGiguere&Kaspi06}
{Faucher-Gigu{\`e}re}, C.-A., \& {Kaspi}, V.~M. 2006, \apj, 643, 332,
  \dodoi{10.1086/501516}

\bibitem[{{Fern{\'a}ndez} \& {Metzger}(2013)}]{Metzger&Fernandez2013}
{Fern{\'a}ndez}, R., \& {Metzger}, B.~D. 2013, \mnras, 435, 502,
  \dodoi{10.1093/mnras/stt1312}

\bibitem[{{Fischer} {et~al.}(2012){Fischer}, {Mart{\'{\i}}nez-Pinedo},
  {Hempel}, \& {Liebend{\"o}rfer}}]{Fischer+12}
{Fischer}, T., {Mart{\'{\i}}nez-Pinedo}, G., {Hempel}, M., \&
  {Liebend{\"o}rfer}, M. 2012, \prd, 85, 083003,
  \dodoi{10.1103/PhysRevD.85.083003}

\bibitem[{Fischer {et~al.}(2010)Fischer, Whitehouse, Mezzacappa, Thielemann, \&
  Liebend{\"o}rfer}]{fischer_protoneutron_2010}
Fischer, T., Whitehouse, S.~C., Mezzacappa, A., Thielemann, F.-K., \&
  Liebend{\"o}rfer, M. 2010, Astron. Astrophys., 517, A80,
  \dodoi{10.1051/0004-6361/200913106}

\bibitem[{Freiburghaus {et~al.}(1999)Freiburghaus, Rosswog, \&
  Thielemann}]{freiburghaus_r-process_1999}
Freiburghaus, C., Rosswog, S., \& Thielemann, F.-K. 1999, Astrophys. J., 525,
  L121, \dodoi{10.1086/312343}

\bibitem[{{Fr{\"o}hlich} {et~al.}(2006){Fr{\"o}hlich}, {Hauser},
  {Liebend{\"o}rfer}, {Mart{\'{\i}}nez-Pinedo}, {Thielemann}, {Bravo},
  {Zinner}, {Hix}, {Langanke}, {Mezzacappa}, \& {Nomoto}}]{Frohlich+06}
{Fr{\"o}hlich}, C., {Hauser}, P., {Liebend{\"o}rfer}, M., {et~al.} 2006, \apj,
  637, 415, \dodoi{10.1086/498224}

\bibitem[{Galeazzi {et~al.}(2013)Galeazzi, Kastaun, Rezzolla, \&
  Font}]{galeazzi_implementation_2013}
Galeazzi, F., Kastaun, W., Rezzolla, L., \& Font, J.~A. 2013, \prd, 88, 064009,
  \dodoi{10.1103/PhysRevD.88.064009}

\bibitem[{Goodale {et~al.}(2003)Goodale, Allen, Lanfermann, Mass{\'o}, Radke,
  Seidel, \& Shalf}]{goodale_cactus_2003}
Goodale, T., Allen, G., Lanfermann, G., {et~al.} 2003, in Vector and {Parallel}
  {Processing} {\textendash} {VECPAR}'2002, 5th {International} {Conference},
  {Lecture} {Notes} in {Computer} {Science} (Berlin: Springer)

\bibitem[{{Gossan} {et~al.}(2020){Gossan}, {Fuller}, \& {Roberts}}]{Gossan+20}
{Gossan}, S.~E., {Fuller}, J., \& {Roberts}, L.~F. 2020, \mnras, 491, 5376,
  \dodoi{10.1093/mnras/stz3243}

\bibitem[{Harten {et~al.}(1983)Harten, Lax, \& van Leer}]{harten_upstream_1983}
Harten, A., Lax, P.~D., \& van Leer, B. 1983, SIAM Review, 25, 35,
  \dodoi{http://dx.doi.org.subzero.lib.uoguelph.ca/10.1137/1025002}

\bibitem[{Hempel \& Schaffner-Bielich(2010)}]{Hempel+10}
Hempel, M., \& Schaffner-Bielich, J. 2010, Nuclear Physics A, 837, 210–254,
  \dodoi{10.1016/j.nuclphysa.2010.02.010}

\bibitem[{{Hoffman} {et~al.}(1997){Hoffman}, {Woosley}, \& {Qian}}]{Hoffman+97}
{Hoffman}, R.~D., {Woosley}, S.~E., \& {Qian}, Y.-Z. 1997, \apj, 482, 951,
  \dodoi{10.1086/304181}

\bibitem[{{Honda} {et~al.}(2006){Honda}, {Aoki}, {Ishimaru}, {Wanajo}, \&
  {Ryan}}]{Honda+06}
{Honda}, S., {Aoki}, W., {Ishimaru}, Y., {Wanajo}, S., \& {Ryan}, S.~G. 2006,
  \apj, 643, 1180, \dodoi{10.1086/503195}

\bibitem[{H\"udepohl {et~al.}(2010)H\"udepohl, M\"uller, Janka, Marek, \&
  Raffelt}]{Hudephol+10}
H\"udepohl, L., M\"uller, B., Janka, H.-T., Marek, A., \& Raffelt, G.~G. 2010,
  Phys. Rev. Lett., 104, 251101, \dodoi{10.1103/PhysRevLett.104.251101}

\bibitem[{{Kajino} {et~al.}(2000){Kajino}, {Otsuki}, {Wanajo}, {Orito}, \&
  {Mathews}}]{Kajino+00}
{Kajino}, T., {Otsuki}, K., {Wanajo}, S., {Orito}, M., \& {Mathews}, G.~J.
  2000, in Few-Body Problems in Physics '99, ed. S.~{Oryu} \& S.~{Kamimura},
  M.~Ishikawa, 80.
\newblock \doarXiv{astro-ph/0006079}

\bibitem[{{Kaplan} {et~al.}(2014){Kaplan}, {Ott}, {O'Connor}, {Kiuchi},
  {Roberts}, \& {Duez}}]{Kaplan+14}
{Kaplan}, J.~D., {Ott}, C.~D., {O'Connor}, E.~P., {et~al.} 2014, \apj, 790, 19,
  \dodoi{10.1088/0004-637X/790/1/19}

\bibitem[{{Lamers} \& {Cassinelli}(1999)}]{Lamers&Cassinelli99}
{Lamers}, H.~J.~G.~L.~M., \& {Cassinelli}, J.~P. 1999, {Introduction to Stellar
  Winds}, ed. {Lamers, H.~J.~G.~L.~M.~\& Cassinelli, J.~P.}

\bibitem[{Lippuner \& Roberts(2015)}]{lippuner_r-process_2015}
Lippuner, J., \& Roberts, L.~F. 2015, Astrophys. J., 815, 82,
  \dodoi{10.1088/0004-637X/815/2/82}

\bibitem[{Löffler {et~al.}(2012)Löffler, Faber, Bentivegna, Bode, Diener,
  Haas, Hinder, Mundim, Ott, Schnetter, \& et~al.}]{Loffler+12}
Löffler, F., Faber, J., Bentivegna, E., {et~al.} 2012, Classical and Quantum
  Gravity, 29, 115001, \dodoi{10.1088/0264-9381/29/11/115001}

\bibitem[{Ma \& Fuller(2019)}]{ma_angular_2019}
Ma, L., \& Fuller, J. 2019, \mnras, 488, 4338, \dodoi{10.1093/mnras/stz2009}

\bibitem[{{Mart{\'{\i}}nez-Pinedo} {et~al.}(2012){Mart{\'{\i}}nez-Pinedo},
  {Fischer}, {Lohs}, \& {Huther}}]{MartinezPinedo+12}
{Mart{\'{\i}}nez-Pinedo}, G., {Fischer}, T., {Lohs}, A., \& {Huther}, L. 2012,
  Physical Review Letters, 109, 251104, \dodoi{10.1103/PhysRevLett.109.251104}

\bibitem[{Mart{\'i}nez-Pinedo {et~al.}(2012)Mart{\'i}nez-Pinedo, Fischer, Lohs,
  \& Huther}]{martinez-pinedo_charged-current_2012}
Mart{\'i}nez-Pinedo, G., Fischer, T., Lohs, A., \& Huther, L. 2012, \prl, 109,
  251104, \dodoi{10.1103/PhysRevLett.109.251104}

\bibitem[{Metzger \& Fern{\'a}ndez(2014)}]{metzger_red_2014}
Metzger, B.~D., \& Fern{\'a}ndez, R. 2014, Mon. Not. R. Astron. Soc., 441,
  3444, \dodoi{10.1093/mnras/stu802}

\bibitem[{{Metzger} {et~al.}(2011{\natexlab{a}}){Metzger}, {Giannios}, \&
  {Horiuchi}}]{Metzger+11b}
{Metzger}, B.~D., {Giannios}, D., \& {Horiuchi}, S. 2011{\natexlab{a}}, \mnras,
  415, 2495, \dodoi{10.1111/j.1365-2966.2011.18873.x}

\bibitem[{{Metzger} {et~al.}(2011{\natexlab{b}}){Metzger}, {Giannios},
  {Thompson}, {Bucciantini}, \& {Quataert}}]{Metzger+11a}
{Metzger}, B.~D., {Giannios}, D., {Thompson}, T.~A., {Bucciantini}, N., \&
  {Quataert}, E. 2011{\natexlab{b}}, \mnras, 413, 2031,
  \dodoi{10.1111/j.1365-2966.2011.18280.x}

\bibitem[{{Metzger} {et~al.}(2007){Metzger}, {Thompson}, \&
  {Quataert}}]{Metzger+07}
{Metzger}, B.~D., {Thompson}, T.~A., \& {Quataert}, E. 2007, \apj, 659, 561,
  \dodoi{10.1086/512059}

\bibitem[{{Metzger} {et~al.}(2008){Metzger}, {Thompson}, \&
  {Quataert}}]{Metzger+08}
---. 2008, \apj, 676, 1130, \dodoi{10.1086/526418}

\bibitem[{{Metzger} {et~al.}(2018){Metzger}, {Thompson}, \&
  {Quataert}}]{Metzger+18}
---. 2018, \apj, 856, 101, \dodoi{10.3847/1538-4357/aab095}

\bibitem[{{Meyer}(2002)}]{Meyer02}
{Meyer}, B.~S. 2002, \prl, 89, 231101, \dodoi{10.1103/PhysRevLett.89.231101}

\bibitem[{{Meyer} \& {Brown}(1997)}]{Meyer&Brown97}
{Meyer}, B.~S., \& {Brown}, J.~S. 1997, \apjs, 112, 199, \dodoi{10.1086/313032}

\bibitem[{{Meyer} {et~al.}(1992){Meyer}, {Mathews}, {Howard}, {Woosley}, \&
  {Hoffman}}]{Meyer+92}
{Meyer}, B.~S., {Mathews}, G.~J., {Howard}, W.~M., {Woosley}, S.~E., \&
  {Hoffman}, R.~D. 1992, \apj, 399, 656, \dodoi{10.1086/171957}

\bibitem[{{M{\"o}sta} {et~al.}(2014){M{\"o}sta}, {Richers}, {Ott}, {Haas},
  {Piro}, {Boydstun}, {Abdikamalov}, {Reisswig}, \& {Schnetter}}]{Mosta+14}
{M{\"o}sta}, P., {Richers}, S., {Ott}, C.~D., {et~al.} 2014, \apjl, 785, L29,
  \dodoi{10.1088/2041-8205/785/2/L29}

\bibitem[{{Nakazato} {et~al.}(2013){Nakazato}, {Sumiyoshi}, {Suzuki}, {Totani},
  {Umeda}, \& {Yamada}}]{Nakazato+13}
{Nakazato}, K., {Sumiyoshi}, K., {Suzuki}, H., {et~al.} 2013, \apjs, 205, 2,
  \dodoi{10.1088/0067-0049/205/1/2}

\bibitem[{Neilsen {et~al.}(2014)Neilsen, Liebling, Anderson, Lehner, O'Connor,
  \& Palenzuela}]{neilsen_magnetized_2014}
Neilsen, D., Liebling, S.~L., Anderson, M., {et~al.} 2014, \prd, 89, 104029,
  \dodoi{10.1103/PhysRevD.89.104029}

\bibitem[{{Otsuki} {et~al.}(2000){Otsuki}, {Tagoshi}, {Kajino}, \&
  {Wanajo}}]{Otsuki+00}
{Otsuki}, K., {Tagoshi}, H., {Kajino}, T., \& {Wanajo}, S.-y. 2000, \apj, 533,
  424, \dodoi{10.1086/308632}

\bibitem[{{Pascal} {et~al.}(2022){Pascal}, {Novak}, \& {Oertel}}]{Pascal+22}
{Pascal}, A., {Novak}, J., \& {Oertel}, M. 2022, arXiv e-prints,
  arXiv:2201.01955.
\newblock \doarXiv{2201.01955}

\bibitem[{Perego {et~al.}(2014)Perego, Rosswog, Cabez{\'o}n, Korobkin,
  K{\"a}ppeli, Arcones, \& Liebend{\"o}rfer}]{perego_neutrino-driven_2014}
Perego, A., Rosswog, S., Cabez{\'o}n, R.~M., {et~al.} 2014, Mon. Not. R.
  Astron. Soc., 443, 3134, \dodoi{10.1093/mnras/stu1352}

\bibitem[{{Perley} {et~al.}(2020){Perley}, {Fremling}, {Sollerman}, {Miller},
  {et~al.}}]{Perley+20}
{Perley}, D.~A., {Fremling}, C., {Sollerman}, J., {Miller}, A.~A., {et~al.}
  2020, \apj, 904, 35, \dodoi{10.3847/1538-4357/abbd98}

\bibitem[{{Perna} {et~al.}(2008){Perna}, {Soria}, {Pooley}, \&
  {Stella}}]{Perna+08}
{Perna}, R., {Soria}, R., {Pooley}, D., \& {Stella}, L. 2008, \mnras, 384,
  1638, \dodoi{10.1111/j.1365-2966.2007.12821.x}

\bibitem[{{Pons} {et~al.}(1999){Pons}, {Reddy}, {Prakash}, {Lattimer}, \&
  {Miralles}}]{Pons+99}
{Pons}, J.~A., {Reddy}, S., {Prakash}, M., {Lattimer}, J.~M., \& {Miralles},
  J.~A. 1999, \apj, 513, 780, \dodoi{10.1086/306889}

\bibitem[{{Qian} \& {Woosley}(1996)}]{Qian&Woosley96}
{Qian}, Y., \& {Woosley}, S.~E. 1996, \apj, 471, 331, \dodoi{10.1086/177973}

\bibitem[{Qian {et~al.}(1993)Qian, Fuller, Mathews, Mayle, Wilson, \&
  Woosley}]{Qian+93}
Qian, Y.-Z., Fuller, G.~M., Mathews, G.~J., {et~al.} 1993, Phys. Rev. Lett.,
  71, 1965, \dodoi{10.1103/PhysRevLett.71.1965}

\bibitem[{{Qian} \& {Wasserburg}(2007)}]{Qian&Wasserburg07}
{Qian}, Y.-Z., \& {Wasserburg}, G.~J. 2007, \physrep, 442, 237,
  \dodoi{10.1016/j.physrep.2007.02.006}

\bibitem[{Qian \& Woosley(1996)}]{qian_nucleosynthesis_1996}
Qian, Y.-Z., \& Woosley, S.~E. 1996, \apj, 471, 331, \dodoi{10.1086/177973}

\bibitem[{Radice {et~al.}(2016)Radice, Galeazzi, Lippuner, Roberts, Ott, \&
  Rezzolla}]{radice_dynamical_2016}
Radice, D., Galeazzi, F., Lippuner, J., {et~al.} 2016, \mnras, 460, 3255,
  \dodoi{10.1093/mnras/stw1227}

\bibitem[{Radice {et~al.}(2018)Radice, Perego, Hotokezaka, Fromm, Bernuzzi, \&
  Roberts}]{radice_binary_2018}
Radice, D., Perego, A., Hotokezaka, K., {et~al.} 2018, \apj, 869, 130,
  \dodoi{10.3847/1538-4357/aaf054}

\bibitem[{{Raynaud} {et~al.}(2020){Raynaud}, {Guilet}, {Janka}, \&
  {Gastine}}]{Raynaud+20}
{Raynaud}, R., {Guilet}, J., {Janka}, H.-T., \& {Gastine}, T. 2020, Science
  Advances, 6, eaay2732, \dodoi{10.1126/sciadv.aay2732}

\bibitem[{{Roberts}(2012)}]{Roberts12}
{Roberts}, L.~F. 2012, \apj, 755, 126, \dodoi{10.1088/0004-637X/755/2/126}

\bibitem[{{Roberts} \& {Reddy}(2017)}]{Roberts&Reddy17}
{Roberts}, L.~F., \& {Reddy}, S. 2017, in Handbook of Supernovae, ed. A.~W.
  {Alsabti} \& P.~{Murdin} (Springer International Publishing AG), 1605,
  \dodoi{10.1007/978-3-319-21846-5\_5}

\bibitem[{Roberts {et~al.}(2012)Roberts, Reddy, \& Shen}]{roberts_medium_2012}
Roberts, L.~F., Reddy, S., \& Shen, G. 2012, \prc, 86, 065803,
  \dodoi{10.1103/PhysRevC.86.065803}

\bibitem[{{Roberts} {et~al.}(2012){Roberts}, {Shen}, {Cirigliano}, {Pons},
  {Reddy}, \& {Woosley}}]{Roberts+12a}
{Roberts}, L.~F., {Shen}, G., {Cirigliano}, V., {et~al.} 2012, Physical Review
  Letters, 108, 061103, \dodoi{10.1103/PhysRevLett.108.061103}

\bibitem[{{Roberts} {et~al.}(2010){Roberts}, {Woosley}, \&
  {Hoffman}}]{Roberts+10}
{Roberts}, L.~F., {Woosley}, S.~E., \& {Hoffman}, R.~D. 2010, \apj, 722, 954,
  \dodoi{10.1088/0004-637X/722/1/954}

\bibitem[{Ruffert {et~al.}(1996)Ruffert, Janka, \&
  Schaefer}]{ruffert_coalescing_1996}
Ruffert, M., Janka, H.-T., \& Schaefer, G. 1996, \aap, 311, 532

\bibitem[{Scheck {et~al.}(2006)Scheck, Kifonidis, Janka, \&
  M{\"u}ller}]{scheck_multidimensional_2006}
Scheck, L., Kifonidis, K., Janka, H.-T., \& M{\"u}ller, E. 2006, \aap, 457,
  963, \dodoi{10.1051/0004-6361:20064855}

\bibitem[{Schnetter {et~al.}(2004)Schnetter, Hawley, \&
  Hawke}]{schnetter_evolutions_2004}
Schnetter, E., Hawley, S.~H., \& Hawke, I. 2004, \cqg, 21, 1465

\bibitem[{Siegel {et~al.}(2013)Siegel, Ciolfi, Harte, \&
  Rezzolla}]{siegel_magnetorotational_2013}
Siegel, D.~M., Ciolfi, R., Harte, A.~I., \& Rezzolla, L. 2013, \prd, 87,
  121302(R), \dodoi{10.1103/PhysRevD.87.121302}

\bibitem[{Siegel {et~al.}(2014)Siegel, Ciolfi, \&
  Rezzolla}]{siegel_magnetically_2014}
Siegel, D.~M., Ciolfi, R., \& Rezzolla, L. 2014, \apjl, 785, L6,
  \dodoi{10.1088/2041-8205/785/1/L6}

\bibitem[{Siegel \& Metzger(2017)}]{siegel_three-dimensional_2017}
Siegel, D.~M., \& Metzger, B.~D. 2017, \prl, 119, 231102,
  \dodoi{10.1103/PhysRevLett.119.231102}

\bibitem[{Siegel \& Metzger(2018)}]{Siegel&Metzger18}
---. 2018, \apj, 858, 52, \dodoi{10.3847/1538-4357/aabaec}

\bibitem[{Siegel \& M{\"o}sta(2018)}]{siegel_grmhd_con2prim_2018}
Siegel, D.~M., \& M{\"o}sta, P. 2018, {GRMHD}\_con2prim: a framework for the
  recovery of primitive variables in general-relativistic magnetohydrodynamics
  (Zenodo), \dodoi{10.5281/zenodo.1213306}

\bibitem[{Siegel {et~al.}(2018)Siegel, M{\"o}sta, Desai, \&
  Wu}]{siegel_recovery_2018}
Siegel, D.~M., M{\"o}sta, P., Desai, D., \& Wu, S. 2018, \apj, 859, 71,
  \dodoi{10.3847/1538-4357/aabcc5}

\bibitem[{{Skinner} {et~al.}(2016){Skinner}, {Burrows}, \&
  {Dolence}}]{Skinner+16}
{Skinner}, M.~A., {Burrows}, A., \& {Dolence}, J.~C. 2016, \apj, 831, 81,
  \dodoi{10.3847/0004-637X/831/1/81}

\bibitem[{{Spite} {et~al.}(2018){Spite}, {Spite}, {Barbuy}, {Bonifacio},
  {Caffau}, \& {Fran{\c{c}}ois}}]{Spite+18}
{Spite}, F., {Spite}, M., {Barbuy}, B., {et~al.} 2018, \aap, 611, A30,
  \dodoi{10.1051/0004-6361/201732096}

\bibitem[{Steiner {et~al.}(2013)Steiner, Hempel, \& Fischer}]{Steiner+13}
Steiner, A.~W., Hempel, M., \& Fischer, T. 2013, \apj, 774, 17,
  \dodoi{10.1088/0004-637x/774/1/17}

\bibitem[{Stergioulas \& Friedman(1995)}]{Stergioulas+95}
Stergioulas, N., \& Friedman, J.~L. 1995, \apj, 444, 306,
  \dodoi{10.1086/175605}

\bibitem[{{Sumiyoshi} {et~al.}(2000){Sumiyoshi}, {Suzuki}, {Otsuki},
  {Terasawa}, \& {Yamada}}]{Sumiyoshi+00}
{Sumiyoshi}, K., {Suzuki}, H., {Otsuki}, K., {Terasawa}, M., \& {Yamada}, S.
  2000, \pasj, 52, 601, \dodoi{10.1093/pasj/52.4.601}

\bibitem[{{Suzuki} \& {Nagataki}(2005)}]{Suzuki&Nagataki05}
{Suzuki}, T.~K., \& {Nagataki}, S. 2005, \apj, 628, 914, \dodoi{10.1086/430847}

\bibitem[{{Takahashi} {et~al.}(1994){Takahashi}, {Witti}, \&
  {Janka}}]{Takahashi+94}
{Takahashi}, K., {Witti}, J., \& {Janka}, H.-T. 1994, \aap, 286, 857

\bibitem[{{Thompson} \& {Duncan}(1993)}]{Thompson&Duncan93}
{Thompson}, C., \& {Duncan}, R.~C. 1993, \apj, 408, 194, \dodoi{10.1086/172580}

\bibitem[{{Thompson}(2003)}]{Thompson03}
{Thompson}, T.~A. 2003, ArXiv Astrophysics e-prints

\bibitem[{{Thompson} {et~al.}(2001){Thompson}, {Burrows}, \&
  {Meyer}}]{Thompson+01}
{Thompson}, T.~A., {Burrows}, A., \& {Meyer}, B.~S. 2001, \apj, 562, 887,
  \dodoi{10.1086/323861}

\bibitem[{{Thompson} {et~al.}(2004){Thompson}, {Chang}, \&
  {Quataert}}]{Thompson+04}
{Thompson}, T.~A., {Chang}, P., \& {Quataert}, E. 2004, \apj, 611, 380,
  \dodoi{10.1086/421969}

\bibitem[{{Thompson} \& {ud-Doula}(2018)}]{Thompson&udDoula18}
{Thompson}, T.~A., \& {ud-Doula}, A. 2018, \mnras, 476, 5502,
  \dodoi{10.1093/mnras/sty480}

\bibitem[{Thornburg(2004)}]{thornburg_black-hole_2004}
Thornburg, J. 2004, \cqg, 21, 3665, \dodoi{10.1088/0264-9381/21/15/004}

\bibitem[{T{\'o}th(2000)}]{toth_nablacdot_2000}
T{\'o}th, G. 2000, \jcph, 161, 605, \dodoi{10.1006/jcph.2000.6519}

\bibitem[{{Vink} \& {Kuiper}(2006)}]{Vink&Kuiper06}
{Vink}, J., \& {Kuiper}, L. 2006, \mnras, 370, L14,
  \dodoi{10.1111/j.1745-3933.2006.00178.x}

\bibitem[{{Vlasov} {et~al.}(2017){Vlasov}, {Metzger}, {Lippuner}, {Roberts}, \&
  {Thompson}}]{Vlasov+17}
{Vlasov}, A.~D., {Metzger}, B.~D., {Lippuner}, J., {Roberts}, L.~F., \&
  {Thompson}, T.~A. 2017, \mnras, 468, 1522, \dodoi{10.1093/mnras/stx478}

\bibitem[{{Vlasov} {et~al.}(2014){Vlasov}, {Metzger}, \&
  {Thompson}}]{Vlasov+14}
{Vlasov}, A.~D., {Metzger}, B.~D., \& {Thompson}, T.~A. 2014, \mnras, 444,
  3537, \dodoi{10.1093/mnras/stu1667}

\bibitem[{von Zeipel(1924)}]{von_zeipel_radiative_1924}
von Zeipel, H. 1924, Mon. Not. Roy. Soc., 84, 665

\bibitem[{{Wanajo}(2013)}]{Wanajo13}
{Wanajo}, S. 2013, \apjl, 770, L22, \dodoi{10.1088/2041-8205/770/2/L22}

\bibitem[{{Winteler} {et~al.}(2012){Winteler}, {K{\"a}ppeli}, {Perego},
  {Arcones}, {Vasset}, {Nishimura}, {Liebend{\"o}rfer}, \&
  {Thielemann}}]{Winteler+12}
{Winteler}, C., {K{\"a}ppeli}, R., {Perego}, A., {et~al.} 2012, \apjl, 750,
  L22, \dodoi{10.1088/2041-8205/750/1/L22}

\bibitem[{{Woosley} \& {Hoffman}(1992)}]{Woosley&Hoffman92}
{Woosley}, S.~E., \& {Hoffman}, R.~D. 1992, \apj, 395, 202,
  \dodoi{10.1086/171644}

\bibitem[{{Woosley} {et~al.}(1994){Woosley}, {Wilson}, {Mathews}, {Hoffman}, \&
  {Meyer}}]{Woosley+94}
{Woosley}, S.~E., {Wilson}, J.~R., {Mathews}, G.~J., {Hoffman}, R.~D., \&
  {Meyer}, B.~S. 1994, \apj, 433, 229, \dodoi{10.1086/174638}

\bibitem[{Yoon \& Langer(2005)}]{yoon_evolution_2005}
Yoon, S.-C., \& Langer, N. 2005, Astron. Astrophys., 435, 967,
  \dodoi{10.1051/0004-6361:20042542}

\end{thebibliography}
\end{document}